\documentclass[preprint]{aastex}
\usepackage{epsf}
\usepackage{graphicx}
\usepackage{threeparttable}

\newcommand{\bfm}[1]{\mathit{\mathbf{#1}}}

\shorttitle{Heating Rate and Emission Line Broadening due to Alfv{\'e}n Wave Turbulence}
\shortauthors{Oran et al.}

\begin{document}

\title{Alfv{\'e}n Wave Turbulence as a Coronal Heating Mechanism: Simultaneously Predicting the Heating Rate and the Wave-Induced Emission Line Broadening}
\author{R. Oran\altaffilmark{1},  E. Landi\altaffilmark{1},  B. van der Holst\altaffilmark{1}, I. V. Sokolov\altaffilmark{1} and  T. I. Gombosi\altaffilmark{1}}

\email{oran@umich.edu}

\altaffiltext{1}{Atmospheric, Oceanic and Atmospheric Sciences, University of Michigan, 2455 Hayward, Ann Arbor, MI, 48109, USA}

\begin{abstract}
In the present work, we test the predictions of the AWSoM model, a global extended-MHD model capable of calculating the propagation and turbulent dissipation of Alfv{\'e}n waves in any magnetic topology, against high resolution spectra of the quiescent off-disk solar corona. Wave dissipation is the only heating mechanism assumed in this model. Combining 3D model results with the CHIANTI atomic database, we were able to create synthetic line-of-sight spectra which include the effects of emission line broadening due to both thermal and wave-related non-thermal motions. To the best of our knowledge this is the first time a global model is used to obtain synthetic non-thermal line broadening. We obtained a steady-state solution driven by a synoptic magnetogram and compared the synthetic spectra with SUMER observations of a quiescent area above the solar west limb extending between 1.04 and 1.34 solar radii at the equator. Both the predicted line widths and the total line fluxes were consistent with the observations for 5 different ions. Using the 3D solution, we were able to locate the region that contributes the most to the emission used for measuring electron properties; we found that region to be a pseudo-streamer, whose modeled electron temperature and density are consistent with the measured ones. We conclude that the turbulent dissipation assumed in the AWSoM model can simultaneously account for the observed heating rate and the non-dissipated wave energy observed in this region.
\end{abstract}

\keywords{magnetohydrodynamics (MHD) - turbulence - Sun:corona - line:profile - method:numerical}

\section{Introduction}
Alfv{\'e}n waves propagating in the solar atmosphere allow for energy transport from the chromosphere into the solar corona and wind, i.e. in the direction opposite to that of heat conduction. These waves have been suggested as a possible energy source for heating the solar corona and accelerating the solar wind \citep{alazraki1971,belcher1971}. This hypothesis is supported by the fact that Alfv{\'e}nic perturbations are ubiquitous in the solar environment, and have been observed in the photosphere, chromosphere, in coronal structures, and in the solar wind at Earth's orbit \citep[c.f.][]{banerjee2011,mcintosh2011}. In addition, the Poynting flux of chromospheric Alfv{\'e}n waves has been shown to be sufficiently large to drive both coronal heating and the solar wind \citep{depontieu2007, mcintosh2012}. In order to account for coronal heating, these waves must undergo some sort of dissipation. Several theoretical models of Alfv{\'e}n wave dissipation have been suggested. These include phase
mixing \citep{Heyvaerts1983}, turbulent cascade \citep{Matthaeus1999}, and resonant absorption \citep{Goossens2011}. However, direct and conclusive observational evidence to support these theories is hard to obtain, due in part to the inherent uncertainty in remote sensing measurements, as we discuss below.\par
Extreme Ultraviolet (EUV) emission by heavy ions provides us with critical tools to study the physical properties and dynamic processes of coronal plasma. Coronal abundances of ions heavier than Helium are low, and therefore these elements do not affect the overall dynamics, but nevertheless their emission in selected spectral lines is routinely observed by spaceborne observatories. While the total line flux depends mainly on the electron density and temperature, the line width is related to the state of the ion responsible for the emission. Specifically, unresolved motions will give rise to Doppler broadening of the spectral line. There are two mechanisms that dominate line broadening in the solar corona: thermal ion motions (due to their finite temperature), and non-thermal ion motions. Non-thermal motions of coronal ions have been suggested to be due to transverse Alfv{\'e}n waves \citep[e.g][]{Hassler1990, Banerjee1998, Doyle1998, Moran2001, Banerjee2009}. Recently, \citet{mcintosh2012} have reported on observational evidence that non-thermal line broadenings are correlated with Alfv{\'e}nic oscillations. Non-thermal line broadening may also be associated with high speed flows taking place in nano-flares \citep{Patsourakos2006}. In this work we study spectral lines formed in the quiet sun, and therefore we do not address the contribution of this mechanism to the line width. Measuring non-thermal mass motions is a difficult endeavor, since both ion temperatures and the non-thermal motions contribute to the observed line width and therefore some assumptions need to be made on the former in order to measure the latter \citep[see][and references therein]{Phillips2008}. \citet{Hahn2012, Hahn2013} studied the observed line broadening in a coronal hole, and found evidence of wave damping. Despite many efforts, direct observational evidence of wave damping in the equatorial corona remain inconclusive. This may be attributed to line-of-sight effects, whereby different spectral lines are actually emitted from different regions.\par 
Several numerical models were aimed at simulating Alfv{\'e}nic perturbations in the solar corona and predicting the observed non-thermal motions. \citet{Ofman1997} generated Alfv{\'e}n waves in a 2.5D resistive magnetohydrodynamics (MHD)  model of an idealized coronal hole. In \citet{Ofman2001} and \citet{Ofman2004} this work was extended to a multi-fluid description in order to directly simulate the motions of the emitting ion species due to a broad band Alfv{\'e}n wave spectrum injected at the base. They directly calculated the resulting line-broadening and found it to agree well with observations. Recently, \citet{Dong2013} have presented results from test-particle simulations showing that a Maxwellian distribution of ion speeds will be broadened when subjected to Alfv{\'e}n waves. They found that the Maxwellian shape is more likely to be preserved during this process when acted on by a wave-spectrum, compared to a monochromatic wave. While these efforts allowed for a detailed description of wave-induced motions, they were restricted to prescribed and idealized magnetic fields. In this work, we wish to extend these efforts to a global model, in which the magnetic field evolves self-consistently with the plasma and wave field, and whose topology can be derived from synoptic maps of the photospheric magnetic field. This allows us to predict EUV line widths and compare them to observations at any location in the lower corona.\par\par 
Several MHD models based on synoptic maps have been developed \citep{usmanov1993,linker1999,mikic1999,roussev2003,Riley2006,cohen2007}. These earlier models employed geometric or empirical heating functions in order to mimic the observed plasma heating and wind acceleration rates, and set their lower boundary at the already hot ($\sim$ 1MK) corona. \citet{lionello2009} and \citet{downs2010} were the first global models which set the inner boundary at the top of the chromosphere. They were able to reproduce the large scale features of the lower corona as observed in full-disk EUV images by introducing different geometric heating functions in coronal holes, streamer belts and active regions. However, global models based on empirical heating functions are limited by the fact that the energy source itself does not evolve self-consistently with the plasma. In addition, \citet{evans2008} found that the Alfv{\'e} speed profiles predicted by such models were less consistent with observations compared to idealized wave-driven models. A self-consistent description of the heating and acceleration in the solar atmosphere can be incorporated into an MHD model by including the effects of Alfv{\'e}n waves, which exchange energy and momentum with the plasma through wave dissipation and wave pressure gradients, respectively.  Alfv{\'e}n waves were first included in a 3D MHD model of the solar corona in \citet{Usmanov2000}, and later in \citet{Usmanov2003}, assuming an ideal dipole magnetic field. These models solved the MHD equations coupled to the wave-kinetic equation for low-frequency Alfv{\'e}n waves of a single polarity, undergoing linear dissipation. A more sophisticated treatment of the dissipation mechanism was implemented in the global model of \citet{vanderholst2010}, which assumed that a Kolmogorov-type non-linear dissipation is taking place in open field line regions, based on the description proposed in \citet{hollweg1986}. This model was validated in \citet{jin2012}, and later extended to include surface Alfv{\'e}n waves in \citet{evans2012}. However, the inner boundary of this model was set at the bottom of the corona with temperatures in the $\sim$1MK range, thus avoiding the problem of forming the corona from the much cooler chromosphere. \par
In this work we use the recently developed Alfv{\'e}n Wave Solar Model (AWSoM) \citep{Sokolov2013,Oran2013}, a global model of the solar atmosphere driven by Alfv{\'e}n wave energy, which is propagated and dissipated in both open and closed magnetic field lines. The model extends from the top of the chromosphere and up to 1-2AU. The interaction of the plasma with the wave field is described by coupling the extended-MHD equations to wave kinetic equations of low-frequency Alfv{\'e}n waves propagating parallel and anti-parallel to the magnetic field. Wave dissipation due to a turbulent cascade is the only heating mechanism assumed in the model. The wave energy in this description represents the time-average of the perturbations due to a turbulent spectrum of Alfv{\'e}n waves. Relating this energy to the non-thermal line broadening, and combining the 3D model results with a spectroscopic database, we are able to calculate synthetic emission line profiles integrated along the entire line-of-sight. The synthetic spectra are used in two ways: First, we compare the synthetic line widths to observations in order to test the accuracy of the model predictions of the Alfv{\'e}n wave amplitude and ion temperatures. Second, the synthetic and observed total line fluxes are compared, in order to test the accuracy of the model predictions of electron density and temperature. In addition, we directly compare the model electron density and temperature to remote measurements based on line intensity ratios. For this purpose, we perform a careful analysis of the emission along the SUMER line of sight as predicted by the model, in order to locate the region that is responsible for the relevant line emission.\par 
This series of independent observational tests allows us to examine whether we can simultaneously account for the coronal plasma heating rate, together with the amount of remaining (non-dissipated) wave energy. Such a comparison provides a vital benchmark for the scenario where coronal heating is due to Alfv{\'e}n wave dissipation. To the best of our knowledge, this is the first time that observed non-thermal mass motions are used to test the heating mechanism in a three-dimensional global model. In the particular case of the AWSoM model, an agreement between the model results and observations would suggest that both the amount of wave energy injected into the system (i.e. the Poynting flux from the chromosphere) and the rate at which the wave energy dissipates at higher altitudes, are consistent with observations.\par 
In order to make meaningful comparisons to observations, we require high quality, high spatial and high spectral resolution data. We selected a set of observations carried out by the Solar Ultraviolet Measurements of Emitted Radiation (SUMER) instrument on board SoHO \citep{Wilhelm1995} during 21-22 November, 1996, in which the SUMER slit was oriented along the solar east-west direction and the SUMER field of view stretched radially from 1.04 to 1.34 solar radii outside the west limb. The AWSoM model was used to create a steady-state simulation for Carrington Rotation 1916 ( 11 Nov. - 9 Dec. 1996), from which we produced synthetic spectra in selected SUMER lines. The radial orientation of the slit allows us to compare predicted and observed quantities as a function of distance from the limb. \par 
This paper is organized as follows. In section \ref{S:nonther} we discuss thermal and non-thermal line broadening of optically thin emission lines. 
In section \ref{S:Model}, we briefly describe the AWSoM model and the numerical simulation for CR1916. The observations used in this study are introduced in section \ref{S:observations}. We describe the method of creating synthetic emission line profiles in section \ref{S:synth}. Section \ref{S:results} reports on the resulting line profiles and their comparison to observations; comparison of the model results to electron density and temperature diagnostics is also shown, and the wave dissipation in the observed region is analyzed. We discuss the results and their implications in section \ref{S:discussion}.
\section{Thermal and Non-thermal Line Broadening}\label{S:nonther}
Unresolved thermal and non-thermal motions of ions will cause emission lines associated with these ions to exhibit Doppler broadening. Outside active regions, the resulting line profile can be approximated by a Gaussian, whose width depends on both the thermal and non-thermal speeds. In the most general case where the non-thermal motions are assumed to be random, the observed full width half maximum (FWHM) of an optically thin emission line will be given by \citep{Phillips2008}: 
\begin{equation}\label{eq:deltalam}
FWHM = \sqrt{\Delta\lambda^2_{inst}+4ln(2) \left(  \frac{\lambda_0}{c}\right)^2\left(\frac{2k_B T_i}{M_i}+v_{nt}^2 \right)},
\end{equation}
where $\Delta \lambda_{inst}$ is the instrumental broadening, $\lambda_0$ is the rest wavelength, $c$ the speed of light, $k_B$ the Boltzmann constant, $T_i$ and $M_i$ are the temperature and atomic mass of ion $i$, respectively, and $v_{nt}$ is the non-thermal speed along the line-of-sight. It is evident from Eq. (\ref{eq:deltalam}) that one cannot determine the separate contributions of thermal and non-thermal motions from the observed FWHM alone. Instead, one must either make some assumption about the ion temperatures or use some model that describes and predicts the magnitude of $v_{nt}$. In this work, we take a different approach, in which we predict both the ion temperatures, $T_i$, and the non-thermal speed, $v_{nt}$ at every location along the line of sight from a global model of the solar atmosphere, and compare the resulting spectra to observations. For this purpose we assume that the non-thermal motions of coronal ions are due to transverse Alfv{\'e}n waves, which cause the ions to move with a velocity equal to the waves velocity perturbation, $\delta\bfm{u}$. In this case the non-thermal speed can be determined according to \citep{Hassler1990, Banerjee1998}:
\begin{equation}\label{eq:vnt}
v_{nt} = \frac{1}{2}\sqrt{<\delta u^2>}|\cos \alpha | ,
\end{equation}
where $<\cdot>$ denotes an average over time scales much larger than the wave period, and $\alpha$ is the angle that the plane perpendicular to the magnetic field makes with the line of sight vector. Eq. (\ref{eq:vnt}) shows that the non-thermal speed is related to the root mean square (rms) of the velocity perturbation rather than to the instantaneous vector. This is due to the fact that line broadening is associated with unresolved motions whose periods are much smaller than the integration time of the detector. The dependence on $\alpha$ reflects the fact that the non-thermal motions due to Alfv{\'e}n waves are inherently anisotropic. The vector $\delta\bfm{u}$ lies in a plane perpendicular to the background magnetic field, and only its component along the line-of-sight contributes to the Doppler broadening of the emission. This dependence on the magnetic field topology is often neglected in works involving coronal holes, but it must be taken into account when considering the equatorial solar corona.\par 
The quantity $<\delta u^2>$ can be calculated from a wave-driven model of the solar corona which describes the evolution of the wave field coupled to an MHD plasma self-consistently. In order to calculate the ion temperatures in detail, one in principle should use a multi-species / multi-fluid MHD description \citep[e.g.][]{Ofman2001, Ofman2004}. Such an approach to a global model of the solar atmosphere is quite involved and is beyond the scope of the present work. However, an extended-MHD description which includes separate electron and proton temperatures might be sufficient, if one assumes that the ions are in thermodynamic equilibrium with the protons. This assumption can be reasonable in the equatorial lower corona due to the high density. Thus, a model that allows the calculation of both the wave amplitude and the proton temperature should be capable of predicting the line broadening under the assumptions we just stated.
\section{Wave-Driven Numerical Simulation}\label{S:Model}
\subsection{AWSoM Model Description}
The Alfv{\'e}n Wave Solar Model (AWSoM) is a global, wave-driven, extended-MHD numerical model starting from the top of the chromosphere and extending into the heliosphere beyond Earth's orbit. The model is based on BATS-R-US, a versatile, massively parallel MHD code, and is implemented within the Space Weather Modeling Framework (SWMF) \citep{toth2012}. The computational domain is based on a non-uniform spherical grid which allows us to treat the sharp gradients in the transition region as well as resolve the heliospheric current sheet. The model employs a unified approach for treating turbulent dissipation in both open and closed magnetic field lines, as presented in \citet{Sokolov2013}. AWSoM was described in detail and validated in \citet{Oran2013}. The model solves the MHD equations coupled to wave-kinetic equations for low-frequency Alfv{\'e}n waves propagating parallel and anti-parallel to the magnetic field. This allows for the exchange of energy and momentum between the wave field and the plasma. Separate pressure equations for electrons and protons allow the inclusion of non-ideal MHD processes such as electron heat conduction, radiative cooling and electron-proton heat exchange. For the sake of brevity we will not repeat the full set of governing equations here and refer the reader to \citet{Sokolov2013} and \citet{Oran2013}. It is worthwhile, however, to briefly discuss the wave dissipation mechanism, since it is the only heating mechanism assumed in the model and it controls the magnitude of the Alfv{\'e}nic perturbations considered in this work. The model assumes that a Poynting flux of Alfv{\'e}n waves is emitted from the top of the chromosphere, with a magnitude proportional to the local magnetic field and constrained by observations (see Table \ref{T:params}). The polarity of the wave emitted from each point on the inner boundary is determined by the direction of the local radial magnetic field. The wave energy densities, $w^+$ and $w^-$, associated with parallel and anti-parallel propagating waves, respectively, are then advected outward along magnetic field lines and dissipated due to a fully developed turbulent cascade \citep{Matthaeus1999}. The energy density dissipation, $Q_w^\pm$, is given by:
\begin{equation}\label{eq:dissipfinal}
Q_w^\pm =\frac{1}{L_\perp\sqrt{\rho}}\sqrt{\max(w^\mp, C_{\rm{ refl}}^2w^\pm)}w^\pm.
\end{equation}
where $\rho$ is the plasma mass density. The dissipation mechanism is controlled by two adjustable parameters: a constant pseudo-reflection coefficient, $C_{\rm{refl}}$, and the transverse correlation length for Alfv{\'e}nic turbulence, $L_{\perp}$, which varies with the width of the magnetic flux tube such that $L_\perp \propto 1/\sqrt{B}$ \citep{hollweg1986}. Note that the dissipation depends on the relative magnitudes of the two wave polarities, leading to different heating rates in open and closed field lines. This unified approach, presented in \citet{Sokolov2013}, ensures that the spatial distribution of coronal heating rates will emerge automatically and self-consistently with the magnetic field topology. A detailed analysis of this approach and its implications can be found in \citet{Oran2013}. The total dissipated wave energy, $Q_w^+ + Q_w^-$, heats both protons and electrons, with 60\% of the heating going into the protons \citep[see][for more details]{breech2009,cranmer2009}. It is important to note that reflections are not directly simulated by the model, rather the pseudo-reflection coefficient $C_{\rm{refl}}$ serves to mimic their effect under the assumption of a fully-developed turbulent cascade. In this approximation any wave energy created by reflections is dissipated locally by the cascade process before it can be carried away by the reflected wave \citep{Matthaeus1999, dmitruk2003, cranmer2007, chandran2009}. Thus in practice there is no need to convert the outgoing wave energy into the opposite polarity, as the wave energy is converted into heat. A less restrictive treatment that includes a self-consistent description of wave reflections in a global model was implemented in \citet{vanderholst2013}. 
\subsection{Relating the Non-thermal Speed to the Modeled Wave Energy}
In the AWSoM model the wave energy evolves under the WKB approximation. The perturbations due to Alfv{\'e}n waves propagating parallel and anti-parallel to the background magnetic field can be conveniently described by the Els\"{a}sser variables, defined as $\bfm{z}_\pm= \delta \bfm{u} \mp \delta \bfm{B}/\sqrt{\mu_0 \rho}$, where $\delta\bfm{u}$ and $\delta\bfm{B}$ are the velocity and magnetic field perturbations, respectively, and $\mu_0$ is the permeability of free space. The wave energy densities can be expressed as $w^\pm=\rho z_\pm^2/4$, while the square of the velocity perturbation can be obtained from:
\begin{equation}\label{eq:zzave}
\delta u^2 = \frac{(\bfm{z}_{+} + \bfm{z}_{-})^2 }{4} = \frac{z_{+}^2 + z_{-}^2 + 2\bfm{z}_{+}\cdot\bfm{z}_{-}}{4}.
\end{equation}
On open field lines, only one wave polarity should dominate if the reflection is negligible so that the product $\bfm{z}_{+}\cdot\bfm{z}_{-}$ will be zero. On closed field lines, opposite wave polarities are injected at the two foot points of the field line, giving rise to counter-propagating waves. However, in the balanced turbulent regime near the top of the closed field lines these perturbations are presumed to be uncorrelated: $<\bfm{z}_{+}\cdot \bfm{z}_{-}>=0$. Thus the last term on the right hand side of Eq. (\ref{eq:zzave}) will drop out in any magnetic topology. The square of the velocity perturbation now becomes:
\begin{equation}\label{eq:zzave2}
\delta u^2 = \frac{z_{+}^2 + z_{-}^2}{4} = \frac{w^+ + w^-}{\rho}.
\end{equation}
Combining Eqs. (\ref{eq:vnt}) and (\ref{eq:zzave2}) we can relate the thermal speed to the wave energies as:
\begin{equation}\label{eq:vntw}
v_{nt} = \frac{1}{2}\sqrt{\frac{w^+ + w^-}{\rho}}| \cos \alpha |.
\end{equation}
Note that under the WKB approximation, the wave energy density is already an average over time scales much larger than the wave period and there is no need for averaging. 
\subsection{Steady-State Simulation for Carrington Rotation 1916}
\begin{figure}[ht]
\epsscale{.80}
\plotone{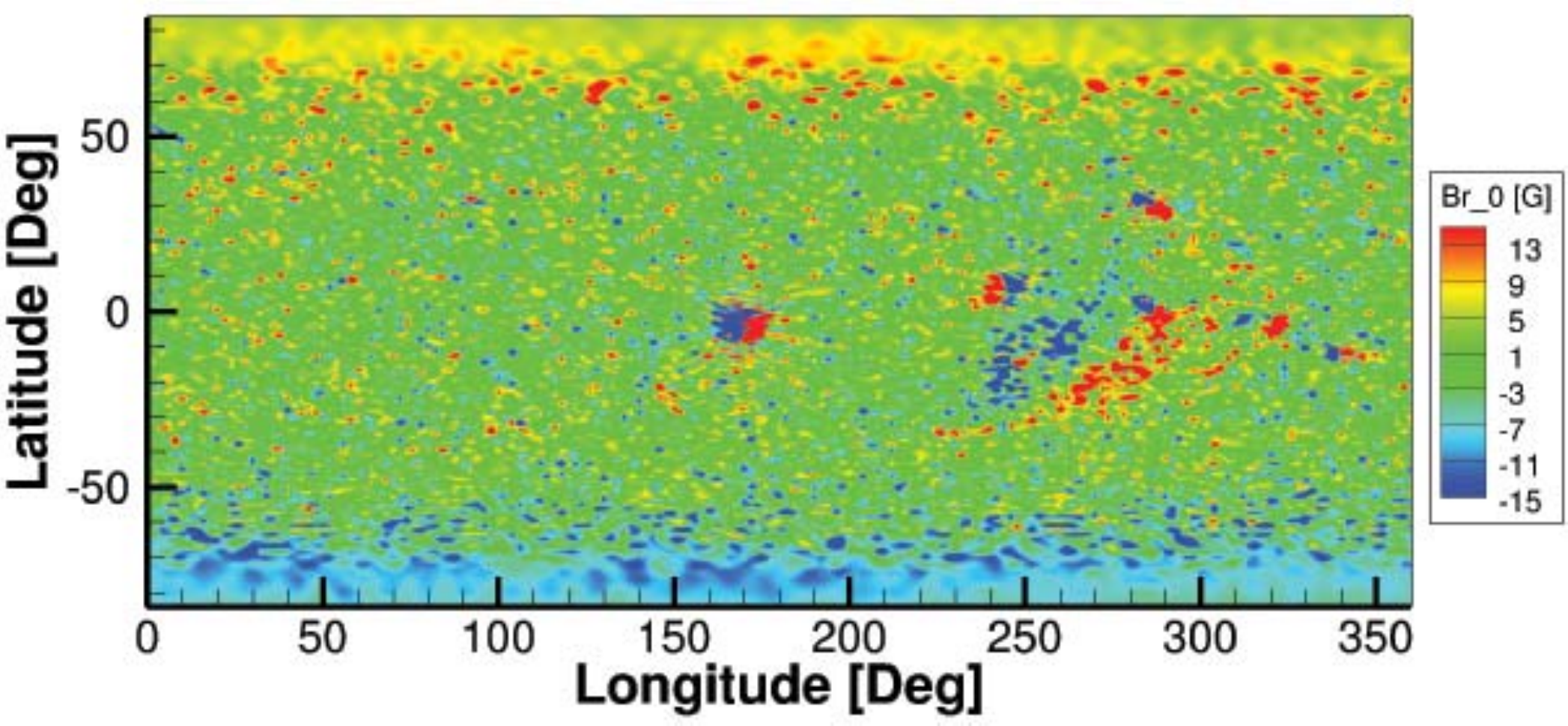} 
\caption{\small \sl Boundary condition for the radial magnetic field for CR1916, obtained from an MDI magnetogram with polar interpolation. Although the magnetic field magnitude can reach up to 2000 G in the vicinity of active regions, the color scale was modified so that the large scale distribution can be seen.\label{F:1916_Br}} 
\end{figure}

In order to produce a realistic steady-state solution for the period during which the SUMER observations were taken, we derive the inner boundary conditions of the model using a synoptic line-of-sight magnetogram of the photospheric radial magnetic field, acquired during Carrington Rotation (CR) 1916 (lasting from 11-Nov-1996 to 9-Dec-1996). The magnetogram was obtained  by the Michelson-Doppler Interferometer (MDI) instrument on board the Solar and Heliospheric Observatory (SoHO) spacecraft \citep{Scherrer1995}.  In order to compensate for the reduced accuracy at polar regions, we use a polar-interpolated synoptic magnetogram, provided by the Solar Oscillations Investigation (SOI) team \citep{Sun2011}.
The resulting radial magnetic field is shown in Figure \ref{F:1916_Br}.\par 
The values used for the model's adjustable parameters and inner boundary conditions for this simulation are listed in Table \ref{T:params}. These values were chosen in accordance with those used for the AWSoM model for a solar minimum case in Oran et al.,(2013), who validated the resulting solution against a myriad of observations, from the lower corona to interplanetary space at 2AU.  The use of the same values for Carrington Rotation 1916, which also took place during solar minimum, is therefore reasonable. Nonetheless, we verify the validity of the global solution used here by comparing model results to full-disk images in Section \ref{S:valid}. A more detailed discussion of these parameters, and their acceptable ranges, can be found in \citet{Oran2013, Sokolov2013}.
\begin{table}[ht]
\centering
\begin{tabular}{| l | l | }
\hline \hline
Input Parameter & Value  \\
\hline\hline
$L_{\perp,0}$ $^*$ & 25 $km$ $^*$\\
\hline
$C_{\rm{refl}}$ & $0.06$ \\
\hline
Poynting flux per unit B $^{**}$ &  $76$ $Wm^{-2}$ $G^{-1}$ \\
\hline
Base electron temperature, $T_e$ & 50,000K  \\
\hline 
Base proton temperature, $T_p$ & 50,000K  \\
\hline 
Base electron density, $n_e$  & $2\times 10^{11}$ $cm^{-3}$  \\
\hline 
Base proton density, $n_p$ &  $2\times 10^{11}$ $cm^{-3}$ \\
\hline \hline 
\end{tabular}
\begin{tablenotes}
\small 
\item $^*$ The correlation length, $L_\perp$, in Eq. (\ref{eq:dissipfinal}) is determined by $L_{\perp} = L_{\perp,0} \sqrt{1[T] / B[T]}$, where $[T]$ denotes a magnetic field measured in units of Tesla.
\item $^{**}$ this value is based on the Hinode observations reported in \citet{depontieu2007}, and corresponds to an rms wave velocity amplitude, $\sqrt{<\delta u^2>} = 12$ km s$^{-1}$ observed at an altitude where the plasma density is $ n_e = n_p = 2\times 10^{10}$ cm$^{-3}$.
\end{tablenotes}
\caption{\small \sl Input parameters and inner boundary values for the AWSoM steady-state simulation for CR1916.}
\label{T:params}
\end{table}

\section{Observations}\label{S:observations}
The observations we used in this work were taken by SUMER instrument on board SoHO on 21--22 November 1996. During this 
time, SoHO was rolled 90 degrees so that the SUMER slit was oriented 
along the East-West direction. The center of the SUMER 4"$\times$300" slit 
was pointed at (0",1160") so that the field of view stretched almost 
radially from 1.04 to 1.34~R$_s$ lying outside the west solar limb at 
the solar equator. The entire 660-1500~\AA\ wavelength range of SUMER 
detector~B was telemetered down; given the particular instrumental 
configuration, this range was divided into 61 sections of 43~\AA, each 
shifted from the previous one by $\approx$13~\AA. Each section was observed 
for 300~s. More details on these observations can be found in \citet{landi2002}.\par 
From the available spectral range, we chose a set of bright and isolated spectral lines (listed in Table \ref{T:lines}), which allow accurate measurements of both line fluxes and line widths up to high altitudes. We note that the very bright O VI doublet at the 1031-1037 \AA ~ range was not selected because these lines are partially formed by radiative scattering from the photosphere, and thus their theoretical FWHM is more complex than given in Eq. (\ref{eq:deltalam}), making them inadequate for our purposes.
\begin{table}[ht]
\centering
\begin{tabular}{| l | l | l | }
\hline \hline
Ion Name & Wavelength [\AA] & R$_{max}$ [$R_S$]  \\
\hline\hline
Fe XII &  1242.0 & 1.275   \\
\hline
S X  & 1196.2 & 1.265  \\
\hline 
Mg IX  & 706.0 & 1.245 \\
\hline 
Na IX & 681.7 & 1.285 \\
\hline 
Ne VIII  & 770.4 & 1.255 \\
\hline \hline 
\end{tabular}
\caption{\small \sl Selected emission lines used in this study. R$_{\rm{max}}$ indicates the highest altitude at which the observed flux is at least 2 times larger than the instrument-scattered flux (see Section \ref{S:stray}).}
\label{T:lines}
\end{table}
\subsection{Data Reduction}
The data were reduced using the standard SUMER software made available by
the SUMER team through the SolarSoft IDL package \citep{freeland1998}; each original frame was flat-fielded,
corrected for geometrical distortions, and aligned with all other frames.
In order to increase the signal-to-noise ratio, the data were averaged
along the slit direction in 30 bins, each 0.01~R$_s$ wide. Spectral
line profiles were fitted with a Gaussian curve removing a linear background.
The resulting count rates were then calibrated using the standard SUMER
calibration also available in SolarSoft. The accuracy of the spectral flux
calibration of SUMER detector~B before June 1998 is $\approx$20\%
\citep[][and references therein]{wilhelm2006}.\par 

\subsection{Scattered Light Evaluation}\label{S:stray}
The micro-roughness of the SUMER optics causes the instrument to scatter the
radiation coming from the solar disk into the detector, even when the instrument is pointing outside the limb. The scattered light forms a ghost 
spectrum of the solar disk at rest wavelength superimposed onto the 
actual spectrum emitted by the region imaged by the SUMER slit.\par  
This ghost spectrum can provide important, though undesired, contributions
to measured line fluxes when the local emission of the Sun is weak;
these contributions need to be evaluated and, when necessary, removed.
Unfortunately, the strength of the ghost spectrum depends on a number
of factors (slit pointing, strength of the disk spectrum etc.) which
make it impossible to devise a procedure to automatically remove it from 
the observations; its estimation needs to be performed on a case-by-case basis.\par 
In the case of the present observations, the almost radial pointing of
the SUMER slit allows us to use the rate of decrease of spectral line
intensities with distance from the limb in order to determine an upper limit on the contributions of the ghost spectrum. Since emission line intensities depend on the square of the electron density, the rapid decrease of the latter with height causes the coronal line intensities to decrease 
by almost two orders of magnitude from the closest to the farthest end
of the slit in the present observation; on the contrary, the scattered light intensity, which is not emitted by the plasma in the observed region, is only reduced by a factor $\lesssim 2$ over the same range.\par 
\citet{landi2007} devised a two-step method to determine an upper limit 
of the scattered light contribution to any spectral line for off-disk
observations stretching over a large range of distances from the limb. 
First, the rate of decrease of the scattered light intensity with height is determined, based on several lines that are not emitted by the corona and whose off-disk intensity is entirely due to scattering. Second, the rate of decrease of scattered light intensity is used to get an upper limit on its contribution to a specific coronal line as follows. We measure the intensity of the coronal line at the location farthest from the limb in the instrument's field of view, and assume that this intensity is entirely due to scattered light. The radial rate of decrease of the scattered light intensity is then normalized to match that coronal line intensity at the same height, giving an upper limit to the scattered light contribution at all other heights. Note that this method actually overestimates the scattered light contribution to coronal lines.\par 
\begin{figure}[h] 
\epsscale{.60}
\plotone{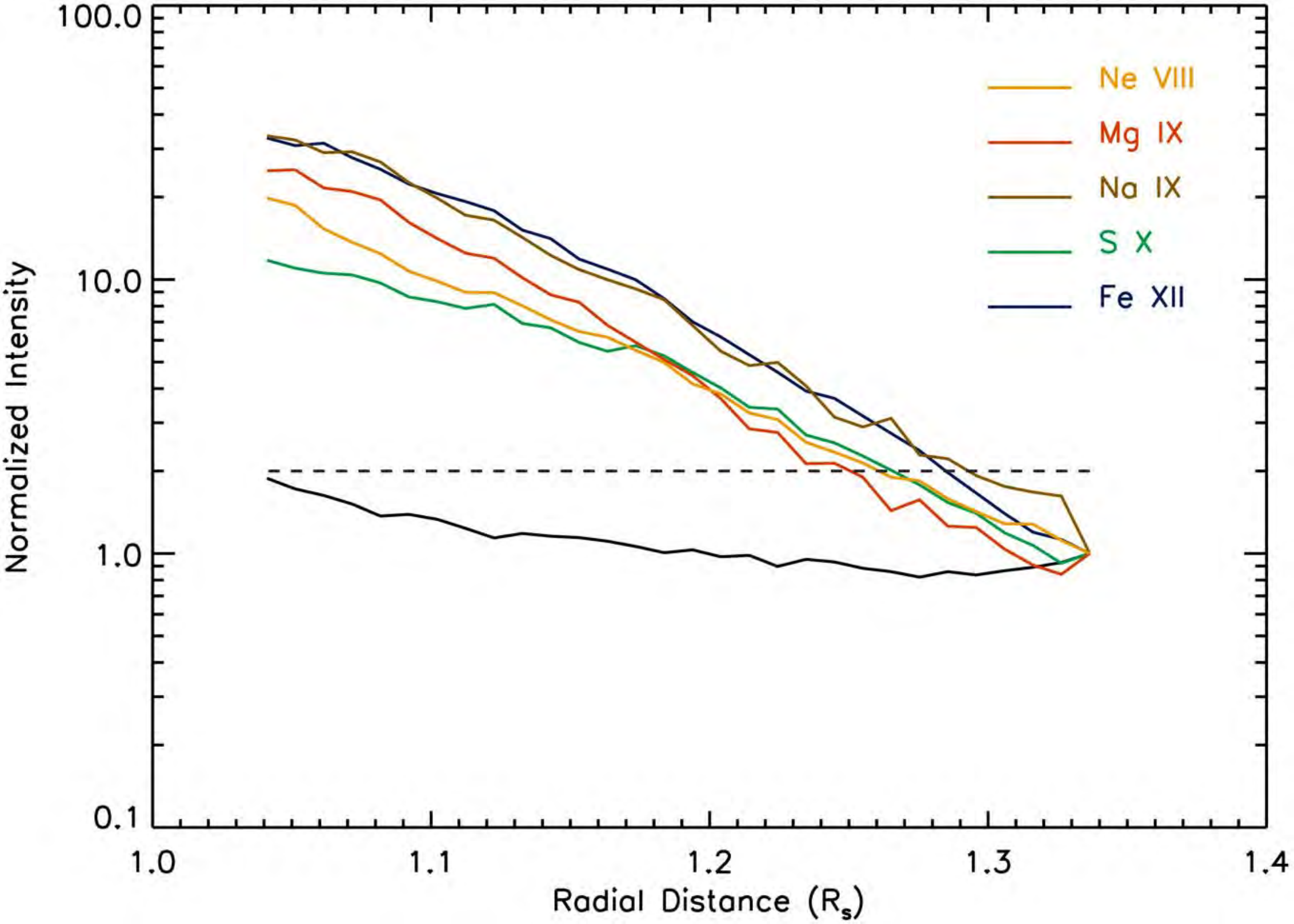} 
\caption{\small \sl Intensity vs. distance for the spectral lines in Table \ref{T:lines}, normalized to the scattered light intensity measured at $r=1.34$ R$_s$ (the farthest point of the SUMER slit). The black solid line shows the averaged scattered light rate of decrease, while the dashed line indicated an intensity level of two times the scattered light intensity at the farthest edge of the slit. \label{F:stray}} 
\end{figure}

To estimate the radial rate of decrease of scattered light intensity, we have used the intensity of the continuum at 1475~\AA,  and of the following lines: 
He I 584~\AA,
C II 1335~\AA,
C III 977~\AA,
O I 1032~\AA, 1304~\AA\ and 1306~\AA, 
O III 835~\AA, and
Si III 1206~\AA. 
These lines and continuum are emitted by the solar chromosphere, so that they are expected to be too weak to be observed at the heights covered by
the SUMER field of view: their observed intensity is entirely due to scattered
light. The rate of decrease of each of these lines and continuum have been normalized to the value of the intensity at the largest distance from the limb and averaged together to provide the final scattered light intensity vs. height curve. This curve appears as the solid black curve in Figure \ref{F:stray}. The normalized intensity vs. height curve for the lines in Table \ref{T:lines} are also shown for comparison. We verified that all of them decreased at a rate much larger than the scattered
light intensity: this suggests that the latter is at best a minor 
contributor to the intensity of each of the lines in Table \ref{T:lines}.
We also determined the maximum heliocentric distance R$_{max}$ below which the scattered light contribution to the coronal line intensity is less than 50\%. We take this 
arbitrary limit as an indication of the range of heights where we can 
safely neglect the scattered light. This height is reported in the third column of Table \ref{T:lines}. We note that all the emission lines considered here possessed a clear Gaussian line shape that could be separated from the background up to distances larger than R$_{max}$.

\section{Synthesizing EUV Emission Line Profiles from 3D Model Results}\label{S:synth}
The synthetic line profiles have been calculated by combining the AWSoM model predictions of the plasma properties and wave energy with the spectral emissivity calculated from the CHIANTI 7.1 atomic database \citep{dere1997,landi2013}. CHIANTI takes into account known line formation mechanisms and is capable of calculating the total emission of a spectral line, given the electron density and temperature. The calculations included in this work were carried out assuming that the plasma is optically thin and in ionization equilibrium. Photo-excitation was neglected as a line formation mechanism.
\subsection{Total Flux of Ion Emission Lines}
The total line emission in a plasma volume, $dV$, having electron temperature $T_e$ and density $N_e$ is given by:
\begin{equation}
\epsilon_{ji} = G_{ji}(N_e, T_e)N_e^2 dV,
\end{equation}
where $G_{ji}(N_e,T_e)$ is the contribution function for a spectral line  associated with an electronic transition from an upper level $j$ to a lower level $i$, defined as:
\begin{equation}\label{eq:GNT}
G_{ji}(N_e,T_e)= A_{ji}\frac{N_j(X^{+m})}{N(X^{+m})}\frac{N(X^{+m})}{N(X)}\frac{N(X)}{N(H)}\frac{N(H)}{N_e}\frac{1}{N_e},
\end{equation}
where $G_{ji}$ is measured in units of photons cm$^{3}$ s$^{-1}$. $X^{+m}$ denotes the ion of the element $X$ at ionization state $+m$. The contribution function also depends on the following quantities:\par 
1. $ N_j(X^{+m}) / N(X^{+m})$ is the relative level population of $X^{+m}$ ions at level $j$, and depends on the electron density and temperature ;\par 
2. $N(X^{+m}) / N(X)$ is the abundance of the ion $X^{+m}$ relative to the abundance of the element $X$, and depends on the electron temperature ;\par  
3. $N(X) / N(H)$ is the abundance of the element X relative
to hydrogen ;\par 
4. $N(H) /N_e$ is the hydrogen abundance relative to the electron density ($\sim$0.83 for a fully ionized plasmas); and\par 
5. $A_{ji}$ is the Einstein coefficient for spontaneous emission for the transition $j\rightarrow i$.\par 
As $T_e$ and $N_e$ are known from the model solution, the contribution function in any computational volume element can be calculated. In this work we used coronal element abundances as given in \citet{feldman1992}, and the latest ionization equilibrium computation available in CHIANTI \citep{landi2013}.\par 
Once the contribution function is calculated at every point along the line-of-sight, the total observed flux in the optically thin limit is given by integrating the emissivity along the line of sight:
\begin{equation}\label{eq:ftot}
F_{tot} = \int \frac{1}{4\pi d^2} G_{ji}(N_e,T_e)N_e^2 dV ,
\end{equation}
where $d$ is the distance of the instrument from the emitting volume $dV$.  $F_{tot}$ is measured in units of photons cm$^{-2}$ s$^{-1}$. This volume integral can be replaced by a line integral by observing that $dV=Adl$, where $A$ is the area observed by the instrument and $dl$ is the path length along the line of sight (LOS). In the case of the present observations, the area covered by the instrument is 4"$\times$1". In order to calculate the LOS integral from the 3D model results, we interpolate $G_{ji}$ and $N_e$ from the AWSoM non-uniform spherical computational grid onto a uniformly spaced set of points along each observed LOS. The spacing used for the interpolation was set to match the finest grid resolution of the model. This procedure ensures that the integration is second-order accurate.
\subsection{Synthetic LOS-integrated Line Profiles}\label{s:unc}
Knowledge of the magnitude of thermal and non-thermal ion motions allows us to calculate a synthetic spectrum, which explicitly includes their effects on the line profile. Thus instead of merely predicting the total flux of an emission line, we can predict the full spectral line profile, to be compared with the observed spectrum.\par 
For each location along the line of sight, the local spectral flux can be calculated by imposing a Gaussian line profile characterized by the predicted total flux, $F_{tot}$, the rest wavelength $\lambda_0$, and line width, $\Delta\lambda$, determined from the ion temperature and the magnitude of non-thermal motions. The spectral flux, measured in units of photons cm$^{-2}$ s$^{-1}$ \AA$^{-1}$, can be written as:
\begin{equation}
F(\lambda) =  F_{tot} \phi(\lambda)  ,
\end{equation}
where $\phi(\lambda)$ is the normalized line profile. In case of a Gaussian line profile, $\phi(\lambda)$ is given by: 
\begin{equation}
\phi(\lambda)=\frac{1}{\sqrt{\pi}\Delta\lambda}exp\left[ -\left(\frac{\lambda - \lambda_0}{\Delta \lambda}\right)\right]  ,
\end{equation}
and the line width, in accordance with Eq. (\ref{eq:deltalam}), can be written as:
\begin{equation}\label{eq:deltanu}
\Delta\lambda = \frac{\lambda_0}{c}\sqrt{\frac{2k_BT_i}{M_i} + v_{nt}^2}.
\end{equation}
The non-thermal speed, $v_{nt}$, can be calculated from the ASWSoM model through Eq. (\ref{eq:vntw}). The emitting region in our case is a three-dimensional non-uniform plasma, where each plasma element along the line-of-sight gives rise to different values of the total flux and the line width. In order to synthesize the line profile from the model, we must perform the line-of-sight integration for each wavelength separately, i.e. we must calculate the spectral flux at the instrument, $F(\lambda)$, given by: 
\begin{equation}\label{eq:spectral_flux}
F(\lambda) = \int \frac{A}{4\pi d^2} \phi(\lambda)  G_{ji}(N_e, T_e) N_e^2 dl .
\end{equation}
The spectral flux is calculated over a wavelength grid identical to the SUMER spectral bins. In order to compare the synthetic spectra with observations, we must also take into account the SUMER instrumental broadening. For this purpose, we convolve the LOS-integrated spectral flux with the wavelength-dependent instrumental broadening for SUMER detector-B, as given by the standard SUMER reduction software available through the SolarSoft package.

\subsection{Uncertainties in Atomic Data and Line Flux Calculations}
Atomic data uncertainties directly affect the line fluxes calculated from the AWSoM simulation results. It is therefore necessary to discuss the accuracy of the data available for the emission lines for which we wish to produce synthetic spectra.
Table \ref{T:lines} lists the five spectral lines that were used for detailed line profile calculations. They were chosen mainly because they are bright and clearly isolated from neighboring lines, so that their profile could be resolved accurately to as large a height as possible.
\subsubsection{Ne VIII 770.4\AA~ and Na IX 681.7\AA~}
These two lines belong to the Li-like iso-electronic sequence, i.e. they possess one bound electron in their outer shell. Their atomic structure is relatively simple and the theoretical calculation of their collisional and radiative rates is expected to be accurate. \citet{landi2002} verified the accuracy of this calculation for all lines belonging to this sequence by comparing the fluxes calculated from CHIANTI to those measured in the 1.04~R$_s$ section of the observations used here.  The authors used the electron density and temperature measured in that section as input to CHIANTI. They found excellent agreement among all lines of the sequence, indicating that the collisional and radiative rates are indeed accurate. However, they found
a systematic factor-2 overestimation of the abundance of all ions of this 
sequence, which they ascribed to inaccuracies in the ionization and
recombination rates used in their work \citep[from][]{mazzotta1998}. 
However, more recent assessments of ionization and recombination rates
made by \citet{Bryans2006, Bryans2009} largely solved this discrepancy, as
shown by \citet{Bryans2009}. Since we
are using ion abundances that take into account the new electron impact
ionization by\citet{Bryans2009}, the fluxes of these two lines
are expected to be reasonably free of atomic physics problems.\\
\subsubsection{Mg IX 706.0\AA~}
The CHIANTI calculation of the flux of this line was found to be in agreement with other lines from the same sequence by \citet{landi2002}; however some problems were found with some other Mg IX line  observed by SUMER, making this ion a candidate for uncertainties in atomic data. However, the radiative and collisional transition rates used in the present work (from CHIANTI 7.1) have been improved from those used by \citet{landi2002}, which used CHIANTI 3 \citep{dere2001}. The new calculations now available in CHIANTI, from \citet{Delzanna2008}, solved the problems so that the atomic data for this ion should be accurate.
\subsubsection{S X 1196.2\AA~}
The atomic data of the S X 1196.2\AA~ line were also benchmarked by\citet{landi2002}, who showed that while all the data in the N-like
iso-electronic sequence were in agreement with each other, they
all indicated a larger plasma electron temperature than the other
sequences, suggesting that improvements in this sequence were 
needed. Subsequent releases of CHIANTI adopted larger and more 
sophisticated calculations for this ion, so that the accuracy
of the predicted flux for S X 1196.2\AA~ should be relatively good. However, this line is emitted by metastable levels in
the ground configuration, and its flux is strongly density sensitive.
Thus, inaccuracies in the predicted electron density may result in large errors in the calculated line flux.
\subsubsection{Fe XII 1242\AA~}
The Fe XII has a complex electronic structure and therefore large atomic models are required to fully describe its wave functions. For example, when EUV lines emitted by this ion are used to measure the electron density, they are known to overestimate it relative
to the values measured from many other ions \citep{Binello2001, young2009, watanabe2009}. The atomic data from \citet{Delzanna2012} in  CHIANTI 7.1 include improved atomic data for this ion, but inaccuracies in the predicted flux of this line may still be expected; in particular, \citet{landi2002} found that 
the atomic data in CHIANTI 3 underestimated the predicted 
flux by $\simeq$30\% while the CHIANTI 7.1 predicted fluxes
are decreased by a factor 1.5-2 compared to Version 3 levels. Thus
we still expect a factor $\approx 2$ underestimation of the total flux of the Fe XII 1242\AA~ line.
\section{Results}\label{S:results}
\subsection{Model Validation for CR1916: EUV Full Disk Images}\label{S:valid}
Comparing observed full disk images to those synthesized from model results allows us to test how well the global, three-dimensional solution, and specifically the temperature and density distributions, can reproduce the observations. Such a comparison also tests the model's prediction of the location and shape of the boundaries between open and closed magnetic field regions, as the coronal holes appear much darker than closed field regions in EUV images.  In the most general case, creating synthetic images requires solving the full radiative transfer through the entire line-of-sight. However, EUV emission lines from the corona and transition region can be treated within the optically thin approximation. This assumption becomes less accurate at the limb, where the optically thin approximation may break down due to the large density along the line of sight. The procedure used to calculate the synthetic images in this work is identical to that presented in \citet{downs2010, Sokolov2013, Oran2013}, and its details will not be repeated here.\par
\begin{figure}[h] 
\plotone{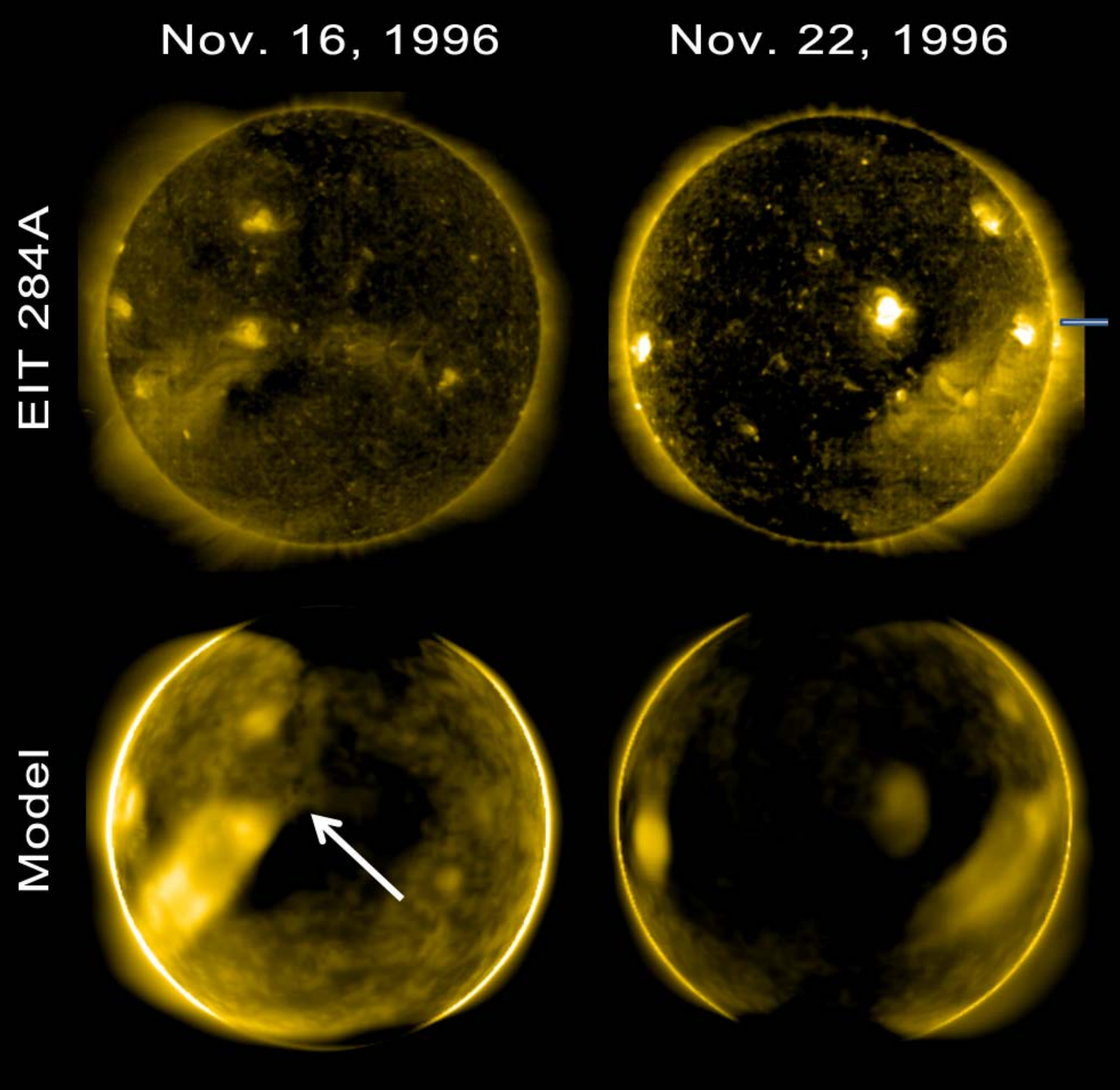} 
\caption{\small \sl SOHO/EIT images vs. synthesized images in the 284\AA $ $ band. Top row shows the observations while the bottom row shows images synthesized from AWSoM. The left column shows images for Nov. 16, 1996 (i.e. a week prior to the observation time), and the white arrow points to the approximate location of the intersection between the SUMER slit and the plane of the sky. The right column shows images for Nov. 22, 1996. The approximate location of the SUMER slit is superimposed on the observed image.\label{F:1916_eit}} 
\end{figure}
We compare our model results for CR1916 to images recorded by the
EUV Imaging Telescope \citep[EIT;][]{Delaboudiniere1995} on board SoHO. In preparing instrument-specific response tables, as well as observed images from the raw data, including calibration, noise reduction and normalization of the photon flux by the exposure time, we used the SolarSoft IDL package.\par
Figure \ref{F:1916_eit} shows observed vs. synthesized images of the $284$~\AA $ $ band, which is dominated by the Fe XV ion, corresponding to an electron temperature of $\sim 2.2$ MK. We present images taken at two different times: the top image shows the solar disk as viewed by SoHO at the time of the SUMER observations, while the bottom figure shows the emission from the solar disk a week earlier, so that the region containing the plane of the sky during the SUMER observation can be viewed close to disk center. As can be seen, the large scale features of the corona, such as coronal hole boundaries and active region locations, are reproduced by the simulation.

\subsection{Comparison of Synthetic and SUMER Spectra}
\begin{figure}[h] 
\epsscale{.6}
\plotone{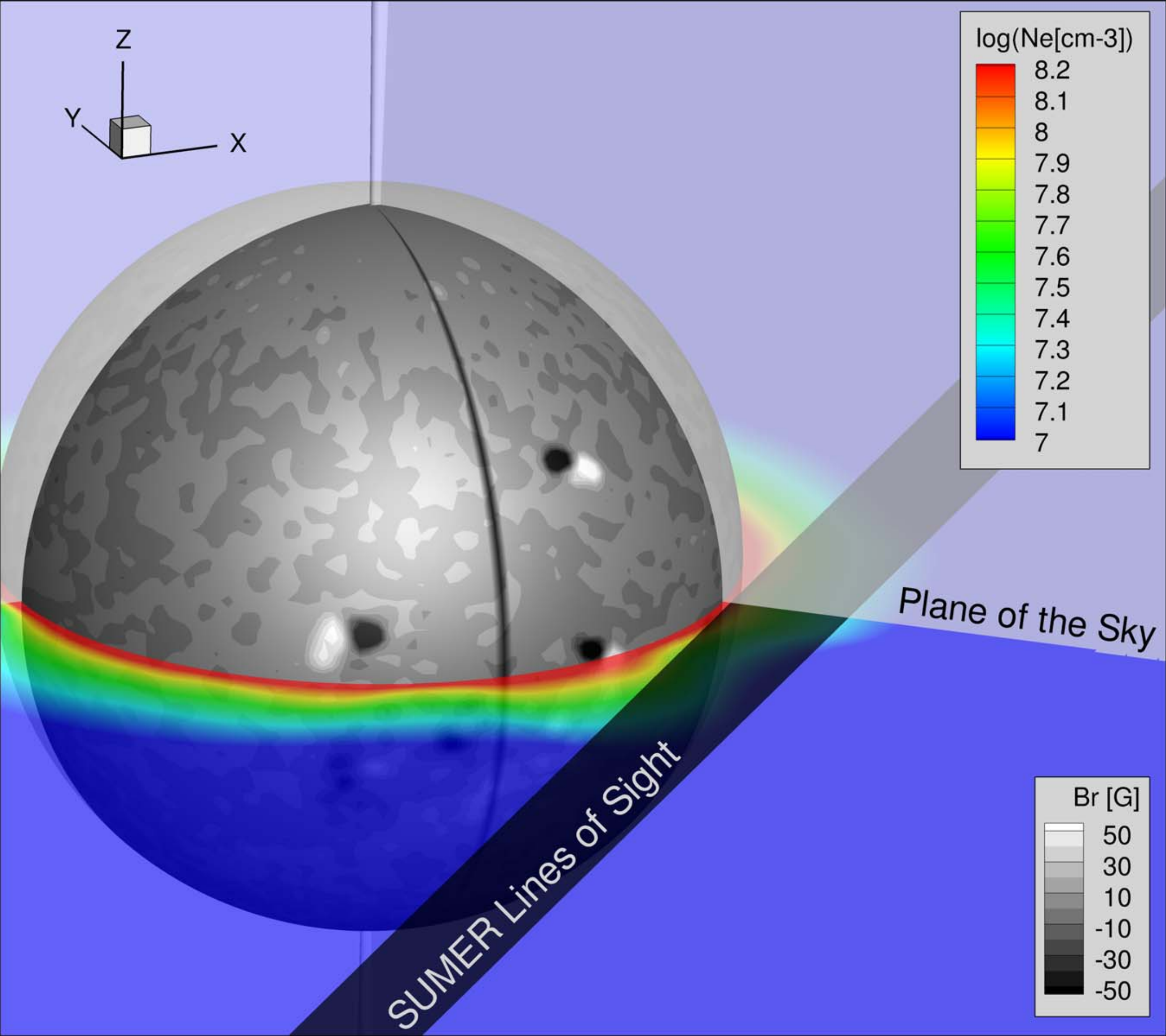} 
\caption{\small \sl 3D model results, location of the plane of the sky, and SUMER lines of sight. The plane containing the SUMER slit is colored by the electron density. The solar surface is colored by the radial magnetic field.\label{F:Slit3D}} 
\end{figure}

In order to perform 3D line-of-sight analysis, we begin with extracting model results, such as electron and proton densities and temperatures, as well as the Alfv{\'e}n waves energy density, along the line of sight to the SUMER observational slit. The geometry of the problem is illustrated in Figure \ref{F:Slit3D}, where the SUMER line-of-sight for the entire slit width is traced within the three-dimensional space of the model solution. The figure shows the solar surface, colored by the radial magnetic field magnitude, the horizontal plane containing the SUMER slit, colored by the electron density, and the plane of the sky for the time of SUMER observations.\par
\begin{figure}[ht]
\epsscale{.9}
\plotone{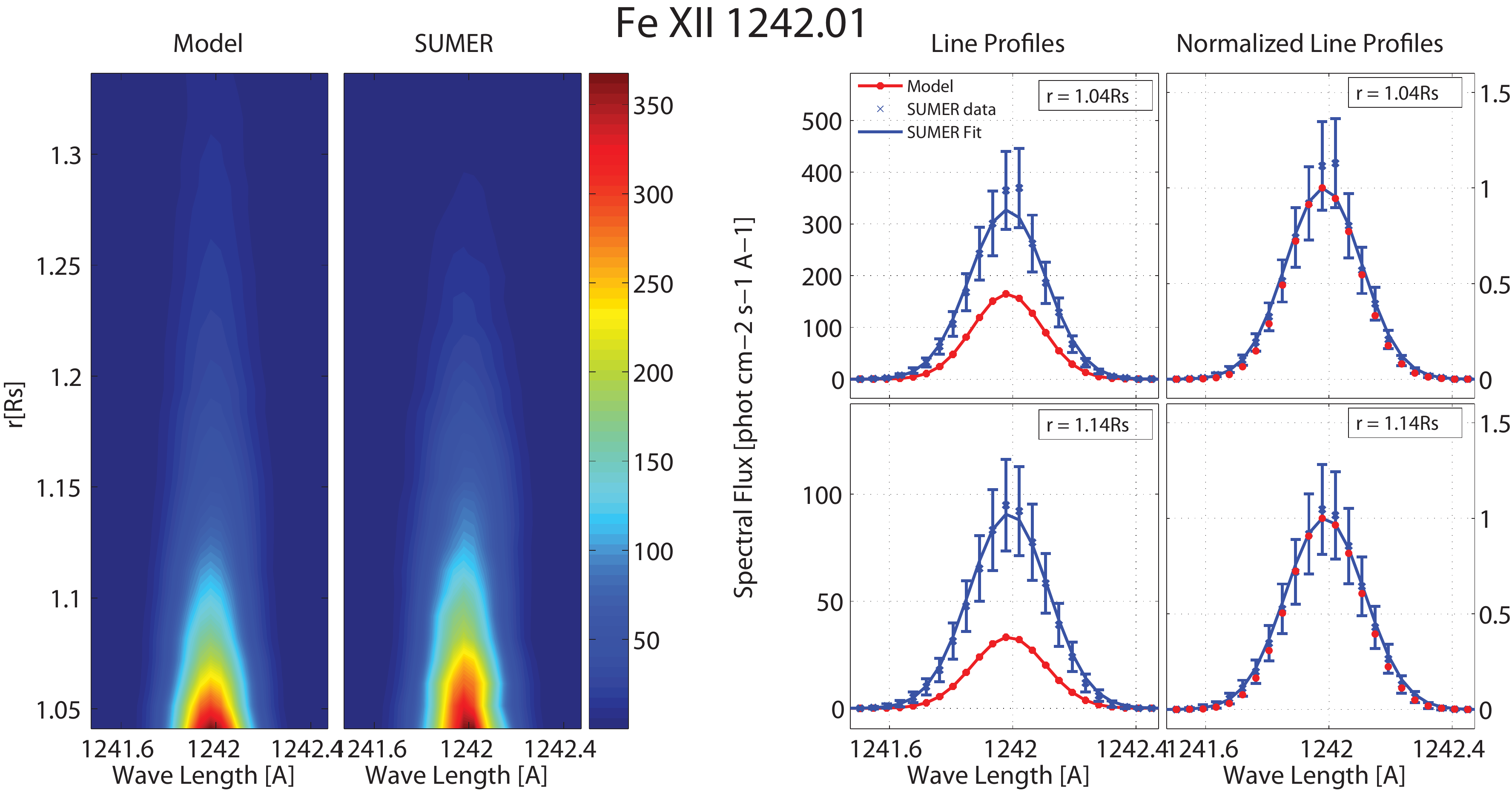} 
\caption{\small \sl Comparison of synthetic and observed spectra for Fe XII 1242~\AA. Left: color plots of synthetic and observed spectra at distances $r=1.04-1.34$R$_s$. Middle: Selected line profiles extracted at $r=1.04$R$_s$ (top) and at $r=1.14$R$_s$ (bottom). Blue symbols with error bars show the SUMER data, the blue solid curve shows the fit to a Gaussian, and the red curve shows the line profile synthesized from the model. Right: Normalized line profiles for the same heights. Curves are color coded in the same way as the middle panels.\label{F:spectra_fe}} 
\end{figure}

\begin{figure}[h]
\plotone{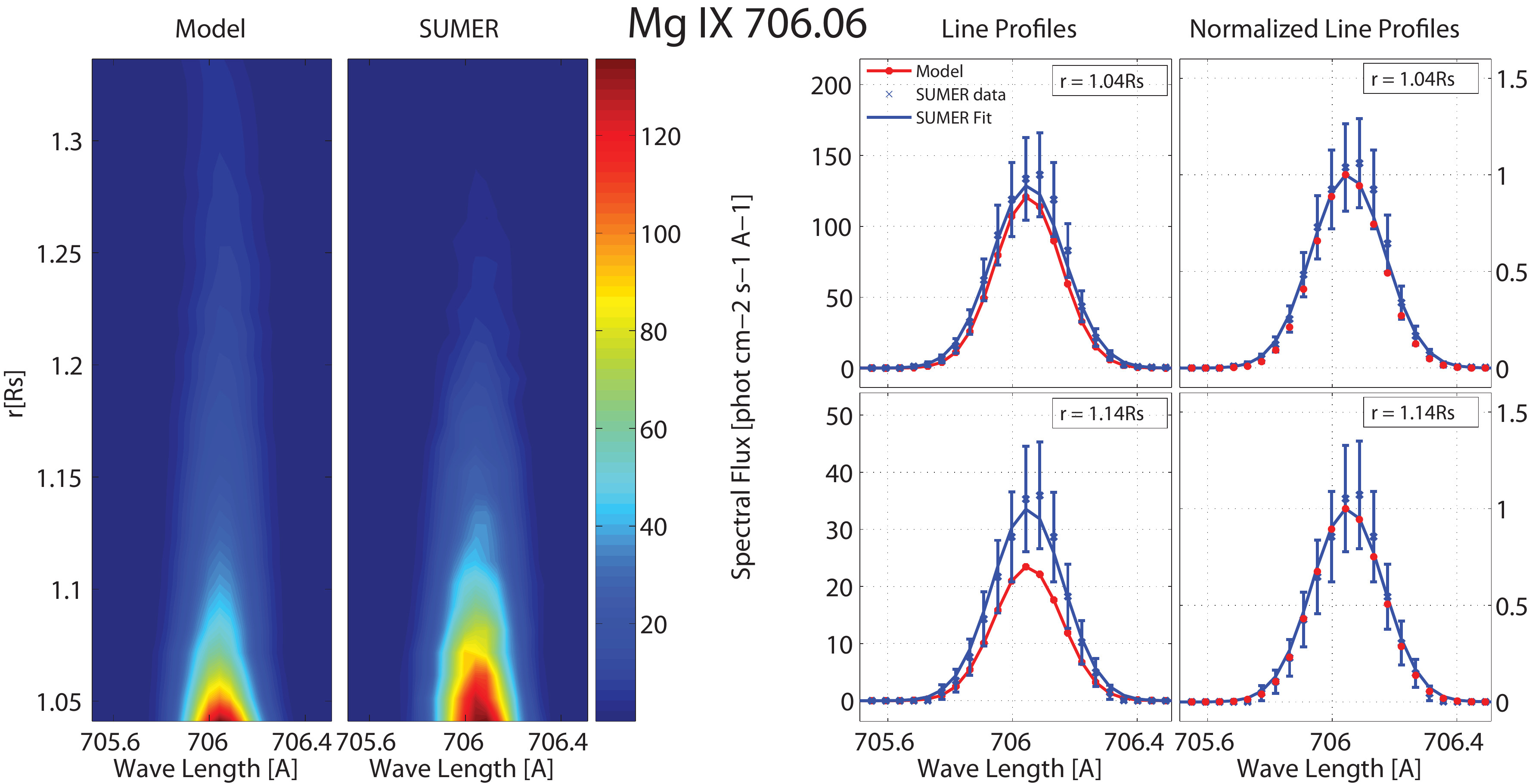} 
\caption{\small \sl Comparison of synthetic and observed spectra for Mg IX 706\AA. See Figure \ref{F:spectra_fe} for the full description.\label{F:spectra_mg}} 
\end{figure}

\begin{figure}[h]
\plotone{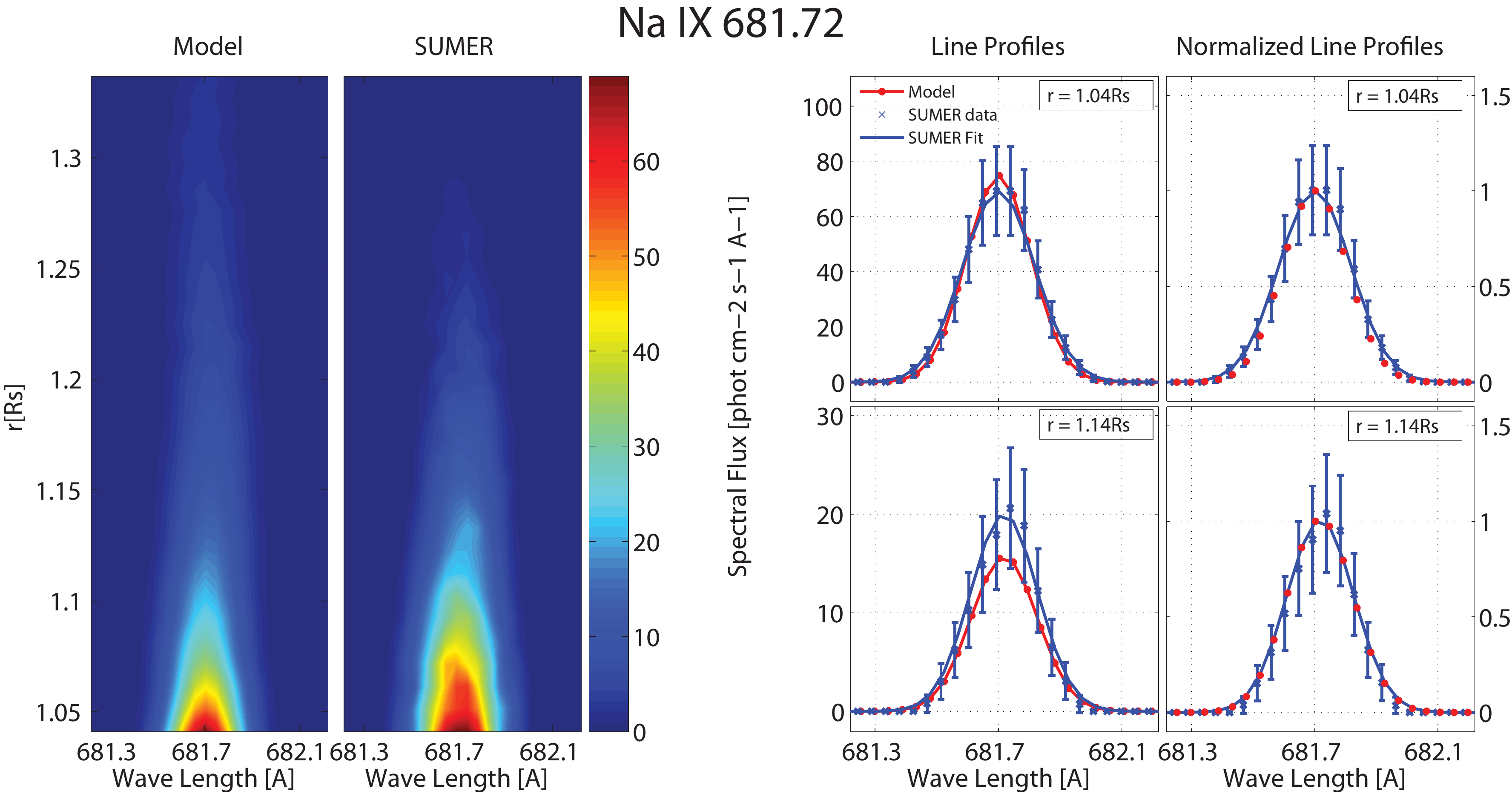} 
\caption{\small \sl Comparison of synthetic and observed spectra for Na IX 681[A]. See Figure \ref{F:spectra_fe} for the full description.\label{F:spectra_na}} 
\end{figure}

\begin{figure}[h]
\plotone{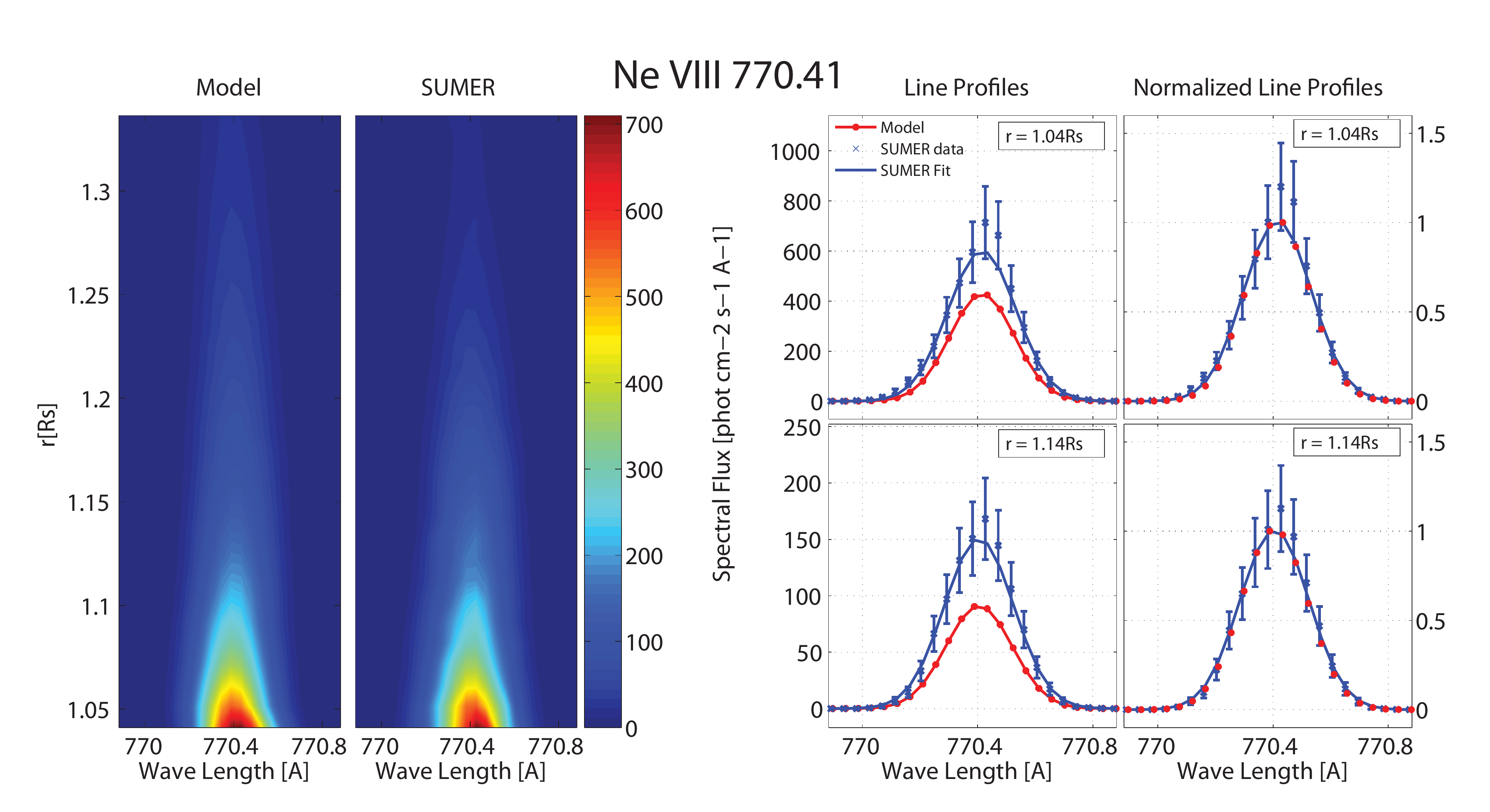} 
\caption{\small \sl Comparison of synthetic and observed spectra for Ne VIII 770\AA. See Figure \ref{F:spectra_fe} for the full description.\label{F:spectra_ne}} 
\end{figure}

\begin{figure}[h]
\plotone{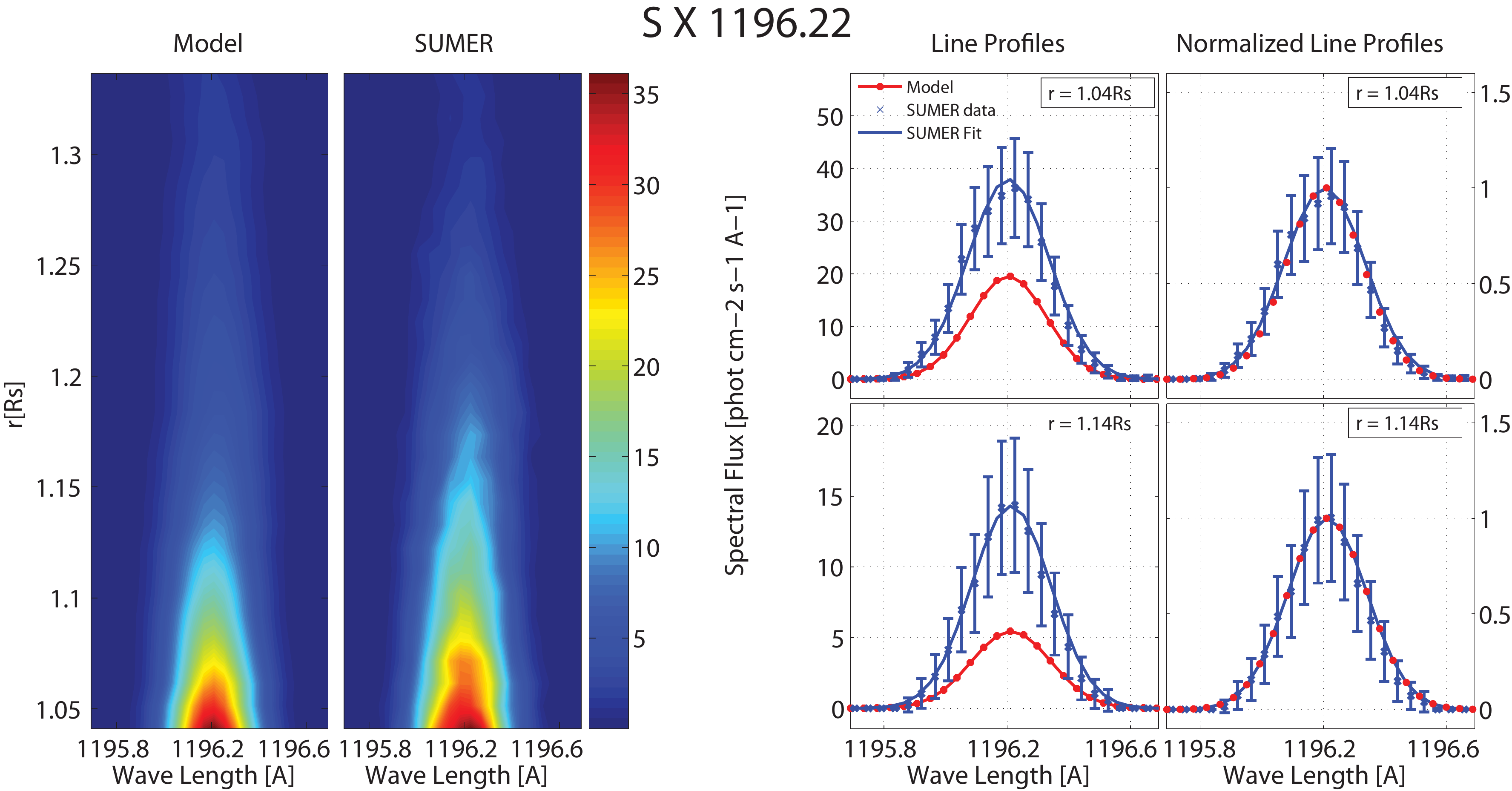} 
\caption{\small \sl Comparison of synthetic and observed spectra for S X 1196\AA. See Figure \ref{F:spectra_fe} for the full description.\label{F:spectra_s}} 
\end{figure}

 
Using the model results and the CHIANTI database, we calculated the spectral flux LOS integral according to Eq. (\ref{eq:spectral_flux}) for each of the lines in Table \ref{T:lines} at each of the 30 radial sections of the SUMER slit. The resulting spectra are compared to the observed spectra in Figures \ref{F:spectra_fe} - \ref{F:spectra_s}. The left panel in each figure shows a contour plot of the synthetic and observed line spectra at all heights covered by the SUMER slit. The middle panel compares the line profile in absolute units at two different distances above the limb: 1.04~R$_s$ and 1.14~R$_s$. The blue symbols and error bars show the observed flux and the associated uncertainty, which takes into account a calibration error of 20\% for SUMER detector-B \citep{wilhelm2006}, and the statistical error in the photon count. The blue curve shows the fit to a Gaussian of the measured flux. The red curve shows the model result. On the right, we show the normalized line profile in each of these heights, using the same color coding as before. The normalized line profile allows us to examine the accuracy of the model prediction of the line width, independent of the absolute value of the predicted total flux. The first thing to notice is that for all lines the observed and predicted line widths are in good agreement at both heights. These results imply that the combination of thermal and non-thermal motions predicted by the AWSoM model is accurate.  The predicted and observed spectral line fluxes are in good agreement for Mg IX and Na IX ions, while the model under predicts their magnitude in the S X, Fe XII and Ne VIII ions. We discuss possible causes of these discrepancies in Section \ref{S:totalflux}.\par 

\clearpage 
\subsection{Comparison of Total Flux vs. Height}\label{S:totalflux}
The total flux predicted by the model depends on the distribution of electron density and temperature along the line of sight. In turn, the radial profiles of the electron density and temperature depend on the heating rate, which in our case is a result of turbulent dissipation of Alfv{\'e}n waves. Thus comparing the radial profiles of the total flux to the observations allows us to verify that the large scale distribution of heating rates predicted by the model give realistic results.\par
\begin{figure}[ht] 
\epsscale{.8}
\plotone{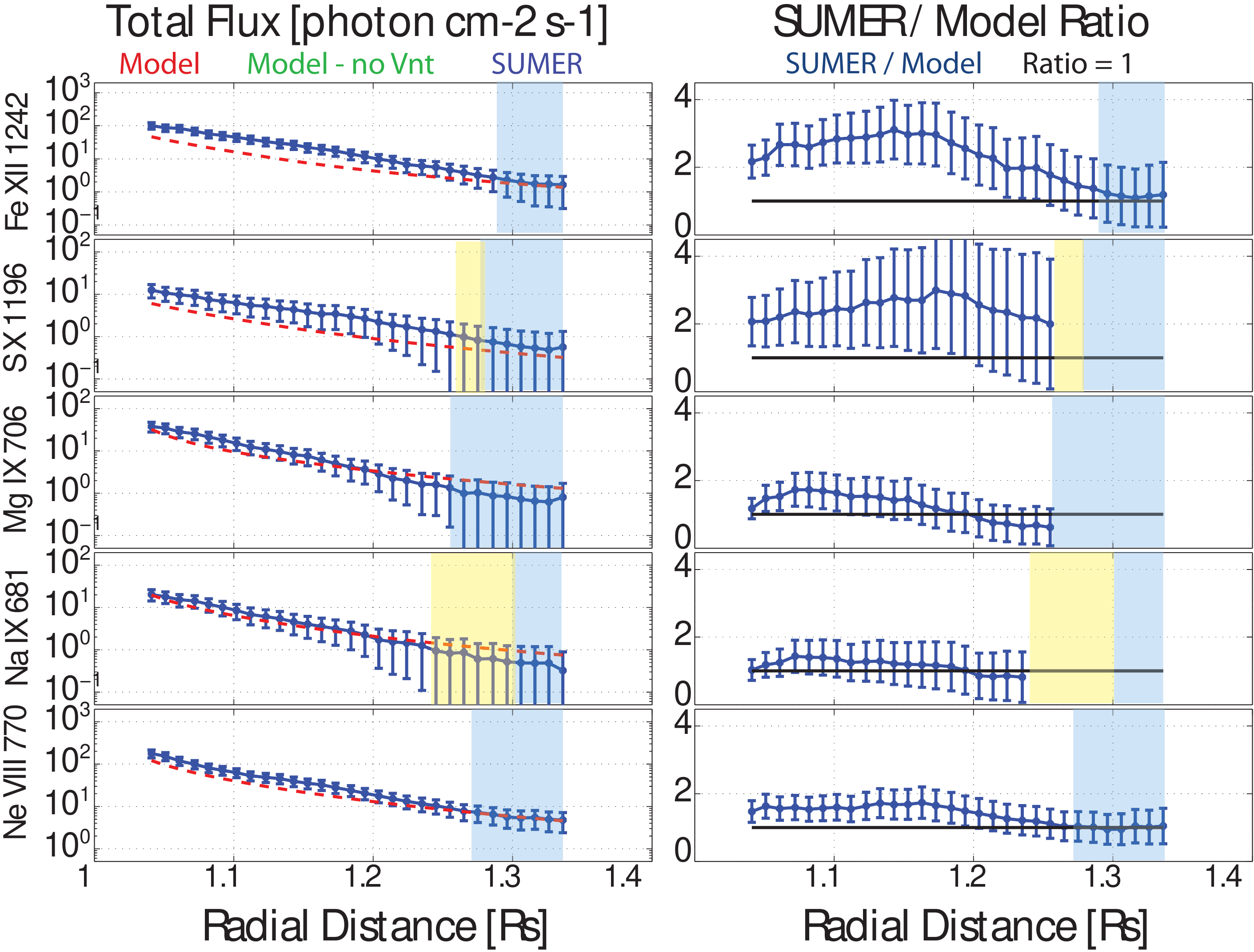} 
\caption{\small \sl Total flux comparison. Left column: Observed (blue) and predicted (red) total fluxes. Right column: Ratio of observed to modeled total fluxes (blue curve). The black curve shows a ratio of 1, for convenience. The regions shaded in orange correspond to heights where the uncertainty in the observed flux becomes larger than the measured value. In this case the uncertainty in the ratio leads to a lower bound that is negative, and therefore, meaningless. The regions shaded in blue corresponds to heights above which the stray light contribution might reach up to 50\% of the observed flux, as reported in Table 2.\label{F:all_flux}} 
\end{figure}

  
Figure \ref{F:all_flux} shows a comparison of the radial profiles of the total flux for all the lines listed in Table \ref{T:lines}. The left panels display the predicted and observed total flux, $F_{tot}$, at all heights covered by the SUMER slit. The panels on the right side of Figure (\ref{F:all_flux}) display the ratio between observed and predicted total 
line fluxes, as a measure to determine the agreement or disagreement
between model and observations. The discrepancies between the model and the observations seem to decrease with radial distance, as all ions show agreement above 1.2 solar radii. However, this decrease is due in part to the increase with height of the uncertainties of the observed fluxes. The regions shaded by an orange color correspond to height where the error in the measured flux is larger than the measured value itself. For these cases, the ratio between predicted and observed total flux becomes meaningless, and these points are excluded from the ratio calculation. The regions shaded in blue correspond to the height above the limb where the scattered light contribution may reach up to 50\% of the observed line flux, as discussed in Section \ref{S:stray}. These heights are summarized in the third column of Table \ref{T:lines}. We next discuss the results for the separate lines in more detail.\par 
\bf{Mg IX and Na IX -} \normalfont The successful comparison for Mg IX and Na IX is very important. Since no atomic physics problems were expected for these lines (see Section \ref{s:unc}), the agreement indicates that the overall temperature and density distributions predicted by the AWSoM model along the line of sight are realistic, although line of sight effects might compensate for local inaccuracies. \par 
\bf{Fe XII - } \normalfont The total flux of the Fe XII line is underestimated, but it is important to note that the factor of 2 to 3 discrepancy we find is similar to the underestimation we expected from this line (see Section~\ref{s:unc}) so that the disagreement could be largely due to atomic data inaccuracies.\par 
\bf{Ne VIII  and S X - } \normalfont The synthetic fluxes for Ne VIII are underestimated by a factor $\approx 1.5$, which is slightly larger than the experimental uncertainties. One possible cause for such a disagreement could be radiation scattering for the Ne VIII line, which we neglected in the present emission calculation. However, \citet{landi2007} showed that radiative scattering is not a significant source of line excitation for Ne VIII below 1.5~R$_s$. The S X line flux is also underestimated by the AWSoM model by a factor $\approx 2$, although the uncertainties on the observed flux are rather large. An overestimation of the electron density along the line of sight might account for part of the disagreement, as the 1196~\AA\ line contribution function, $G(N_e,T_e)$, defined by Eq. (\ref{eq:GNT}), decreases as the density increases beyond $N_e = 10^8$~cm$^{-3}$. However, the discrepancy between the predicted and observed fluxes of both Ne VIII and S X could be due to an inaccurate estimation of their abundances. Coronal element abundances are affected by the fractionation processes active in the corona known as the ``FIP effect'' \citep[][and references therein]{feldman2000}. It has been observed that the abundances ratio of elements with a low ($<10$~eV) First Ionization Potential (FIP) to elements with a high FIP is larger in the corona compared to the photosphere, by a factor known as the ``FIP bias''. The coronal abundances used in the present calculation \citep[from][]{feldman1992} adopt a FIP bias of 4. However, the FIP bias of S is not known with accuracy: \citet{feldman1992} report a FIP bias of 1.15, while, for example, \citet{feldman1998} indicate a FIP bias between 1.2 and 2.0, which is large enough to account for the disagreement we find. The FIP bias of Ne has never been measured, since the photospheric abundance of Ne is unknown. Theoretical models of the FIP effect suggest that Ne is also affected by this process \citep[][and references therein]{laming2012}, so that the absolute abundance of this element in the corona is also subject to uncertainty.
These uncertainties might be causing the discrepancies we find in the total fluxes of these two lines.
\subsection{Comparison of Line Width vs. Height}
The comparison of the radial variation of the line width in the synthetic spectra to that found in the observations allows us to determine how well the predicted plasma and wave properties are able to account for the observed line broadening in the inner ($1.04-1.34$R$_s$) part of the equatorial solar atmosphere.\par
\begin{figure}[ht] 
\plotone{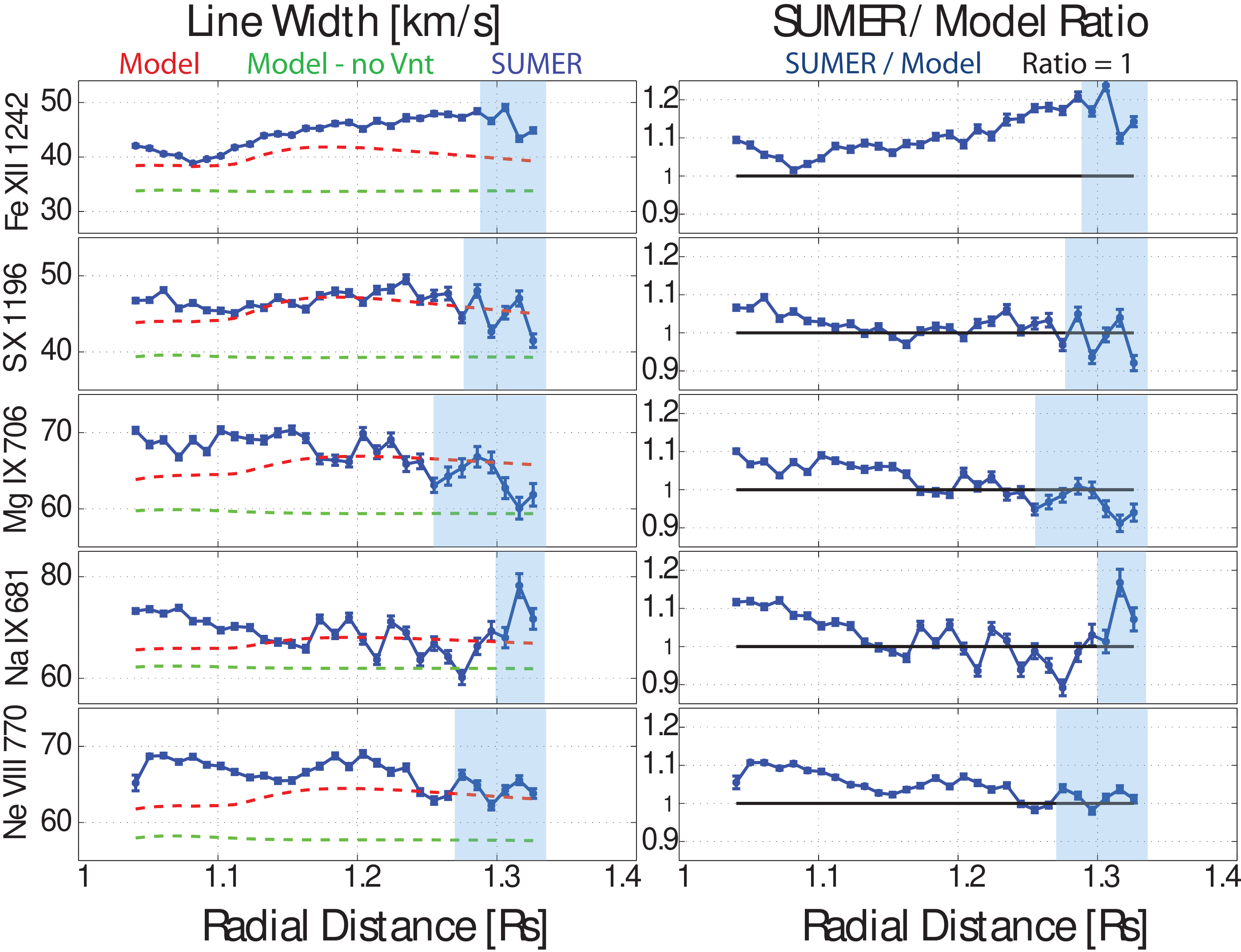} 
\caption{\small \sl Model - SUMER comparison of FWHM.  Blue curves with error bars show the measured FWHM. Red curved show the AWSoM predicted FWHM. Green curves show the AWSoM prediction if the non thermal speed is not be taken into account.\label{F:all_FWHM}} 
\end{figure}

 
Figure \ref{F:all_FWHM} compares the radial profiles of the synthetic and observed line widths for each of the spectral lines in Table \ref{T:lines}. The regions where the scattered light flux may contribute up to 50\% to the line flux are shaded in blue. These radial distances are reported in Table \ref{T:lines}. The panels on the left hand side show the model and observed width cast in units of speed using Width(km s$^{-1}$) = $(\Delta \lambda / \lambda_0)c$, where $c$ is the speed of light in km s$^{-1}$. This quantity is often referred to as the effective speed. The blue curve with error bars shows the observations, while the red dashed line shows the model results. In order to examine the relative contribution from the thermal and non-thermal speeds, we repeated the calculation of the line widths while ignoring the non-thermal speed as a line broadening mechanism. The results are shown as the green curves on the left panels. The panels on the right hand side show the ratio of the observed to synthetic line width (blue curve). The solid black line denotes a ratio of one, i.e. a perfect agreement. The first thing we note is that the ratios for all lines are all very close to unity, with a discrepancy of less than 10\% at most heights. This implies that the combination of ion temperatures and non-thermal speeds predicted by AWSoM can produce synthetic line widths whose magnitudes are very close to the observed ones, at least in the case of the lower equatorial corona. As in the case of the total flux comparisons, we note that line-of-sight effects may compensate for any local inaccuracy in the AWSoM prediction. The removal of the non-thermal speed from the calculation of the synthetic profiles greatly reduces the agreement between the model and the observations. This implies that the non-thermal motions induced by the waves are necessary for predicting line widths which are consistent with observations. While the line width due to thermal motions alone does not change considerably with radial distance, the total line width which includes the wave-induced motions shows a clear radial dependence. This dependence is due in part to the effects of the magnetic topology, as we will discuss in Section \ref{S:ducurve}. This comparison also sheds some light on the validity of our assumption that all the ions have the same temperature. Since the spectral lines considered here are emitted by different elements, the thermal contribution to the line width is different for each of them, while the non-thermal contribution is the same. The simultaneous agreement of the predicted and observed widths for several ions make is less probable that their temperatures do in fact differ from one another. We note that the agreement between the synthetic and observed line widths decreases as the height above the limb increases for the case of Fe XII. This discrepancy may be due to the uncertainty in the observations, but it is also possible that our assumption that this ion, which has the largest mass, has the same speed as the protons breaks down at higher altitudes, where the density has already fallen off considerably and the plasma becomes collisionless.

\subsection{Comparison of Electron Properties}
In the previous sections, we showed that the modeled wave amplitude is consistent with observed line-widths of several different ions, suggesting that the model correctly predicts the amount of wave energy propagating in the corona. To complete this discussion, we wish to verify that the observed coronal heating rate, which depends on the wave dissipation rate, is also reproduced. Since the heating rate impacts the electron density and temperature, comparing the modeled and measured electron properties along the SUMER slit serves as an independent check on the dissipation mechanism assumed in the model. \citet{Oran2013} found that the AWSoM model's prediction of electron properties in a polar coronal hole during solar minimum were in good agreement with measurements. The simple geometry of the coronal hole allowed the authors to compare the line-of-sight measurements to model results extracted along the coronal hole axis. However, in the present case of observations of the equatorial quiet corona, which exhibits a more complex magnetic topology, it becomes less clear which region along the line-of-sight should be compared to the measurements. We therefore adopt a more detailed approach, one that takes into account the variable emission from different magnetic structures crossing the line-of-sight. 
\subsubsection{Overcoming Line-of-Sight Effects: 3D Emission Analysis}\label{S:maxemiss}
The advantage of a three-dimensional model is that it enables us, when combined with the CHIANTI atomic database, to calculate the relative contribution of each emitting volume along the line-of-sight to the total observed emission using the calculations presented in Section \ref{S:synth}. This allows us to assess the amount of contamination to a given coronal structure from emission in the background and foreground, as well as guide us in the interpretation of diagnostic results. We here concentrate on electron density and temperature diagnostics; the electron density along the SUMER slit was measured using the line flux ratio of S X 1196\AA~ and S X 1212\AA, while the electron temperature was measured using the line flux ratio of Mg IX 706\AA~ and Mg IX 749\AA. If a single, well-defined magnetic structure can be identified as a major source of the emission in these lines, then the corresponding modeled quantity in that structure may be compared to the measurement results. We must also require that the relative contribution of this region to the total emission is the same for each of the lines used in the flux ratio calculation. In this way, the ratio of the line fluxes integrated over the selected region will be equal to the ratio of line fluxes integrated over the entire line-of-sight, making the comparison to the observations appropriate. The procedure is somewhat different in case of electron density and temperature measurements, and we discuss these separately. \par
\subsubsection{Region of Maximum Emission for Electron Density Measurements}
\begin{figure}[ht] 
\plotone{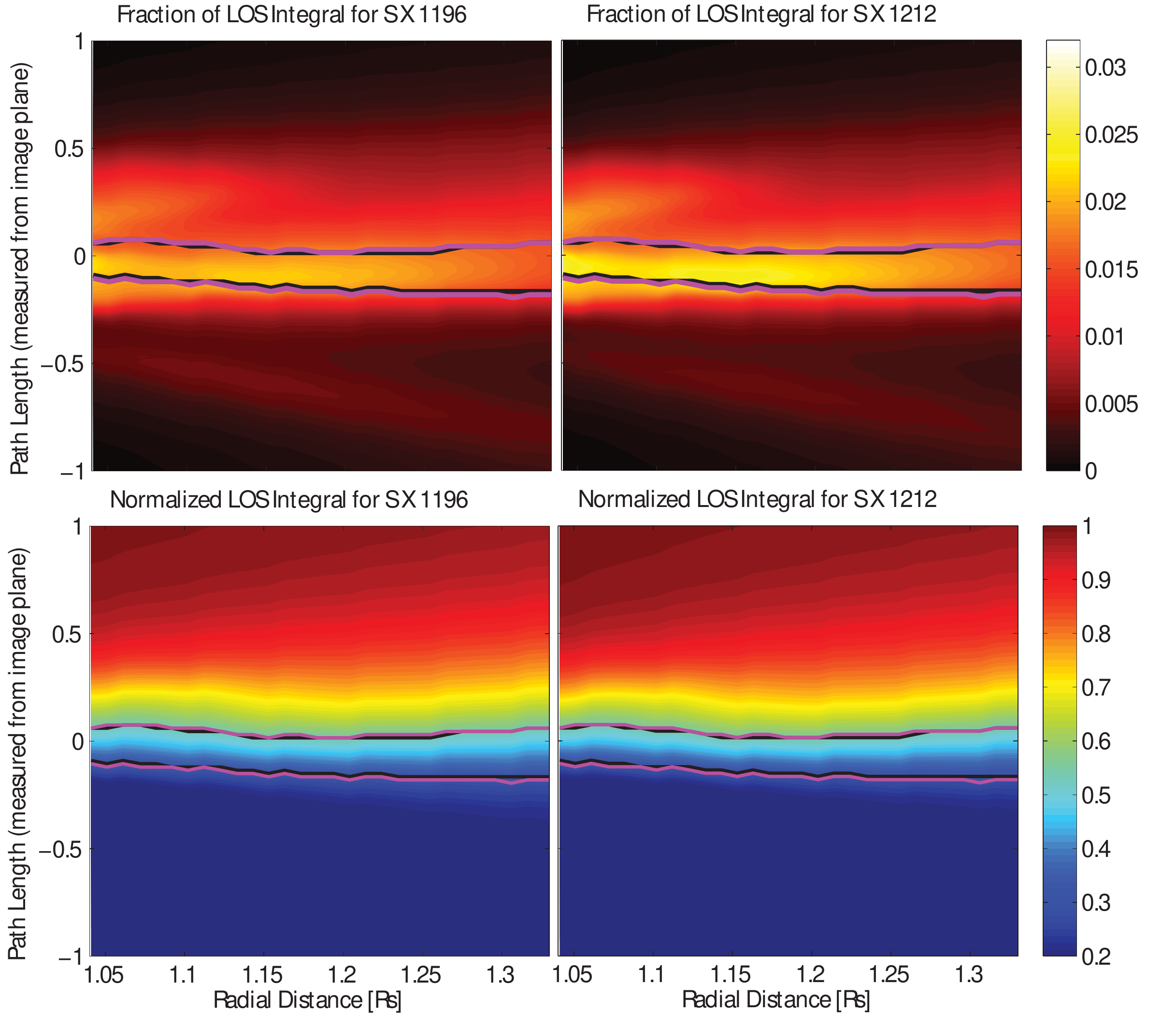} 
\caption{\small \sl Emissivity line of sight integral for  S X 1196\AA (left column) and S X 1212\AA (right column). The top row shows the fractional contribution to the total line of sight integral, along all 30 SUMER lines of sight used in this study. The bottom row shows the cumulative contribution to the LOS integral. The purple curves represent the ranges along the LOS that account for 24\% of the total emission of S X 1196\AA~, while the black curve represents the region that accounts for 24\% of the total emission in the S X 1212\AA~ line.\label{F:SXlines} } 
\end{figure}

The electron density along the SUMER slit was obtained from the line flux ratio of the S X 1196\AA~ and S X 1212\AA~ lines. Figure \ref{F:SXlines} shows the relative contribution of each location along the line of sight to the total emission, calculated using the AWSoM results and the CHIANTI database. The top row shows the fractional contribution to the total emission along each of the lines of sight. The bottom panels show the cumulative normalized LOS integral of the emission for these lines, which ranges from 0 to 1 (corresponding to the two edges of the line of sight). It can be seen that for both lines, the strongest emission comes from a narrow region around the plane of the sky (where we set the path length to 0). At lower altitudes, there is a significant contribution coming from an additional region behind the central region. We have found that $\sim 24\%$ of the total emission of both lines comes from a region that is less than $0.2$R$_s$ wide, marked by the black and purple curves. The black curves show the bounds of the $24\%$ region for S X 1196\AA~, while the purple curves show the same for S X 1212\AA~. Since the two regions more or less overlap, the density modeled in this region is suitable for comparison with the density measurement.\par
\begin{figure}[ht] 
\epsscale{.7}
\plotone{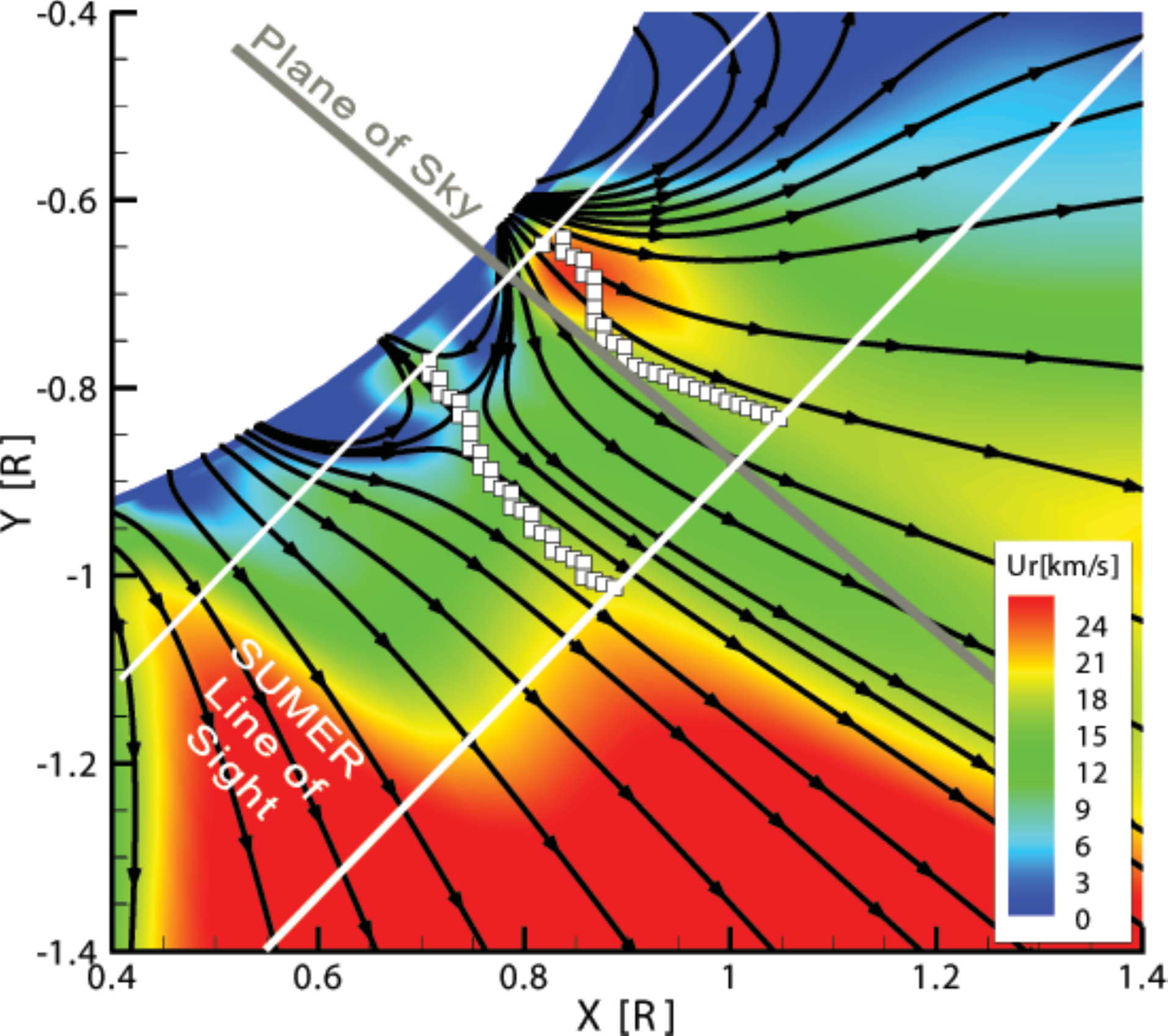} 
\caption{\small \sl Location of maximum emission for the S X 1196~\AA. The color contours show the radial flow speed in the equatorial plane containing the SUMER lines-of-sight (marked by the two white lines). The gray line denotes the plane of the sky for the day the observations. Black stream lines show the magnetic field. White squares show the bounds of line of sight segments corresponding to the purple curves in Figure \ref{F:SXlines}. \label{F:S_10_tec}}
\end{figure}

 
We next locate this region in the model's three-dimensional magnetic topology. Figure \ref{F:S_10_tec} shows the MHD solution in an equatorial plane. Color contours show the radial speed while black curves show the magnetic field. The boundaries of the $24\%$ region for S X 1196\AA~ are marked by the white squares (corresponding to the purple curves in Figure \ref{F:SXlines}). Interestingly enough, we see that a large part of the emission is coming from a distinct magnetic structure of a pseudo-streamer, i.e. a loop structure topped by open field lines of a single polarity. The flow speed above the streamer is slower than the surrounding regions.

\subsubsection{Region of Maximum Emission for Electron Temperature Measurements}
\begin{figure}[ht] 
\plotone{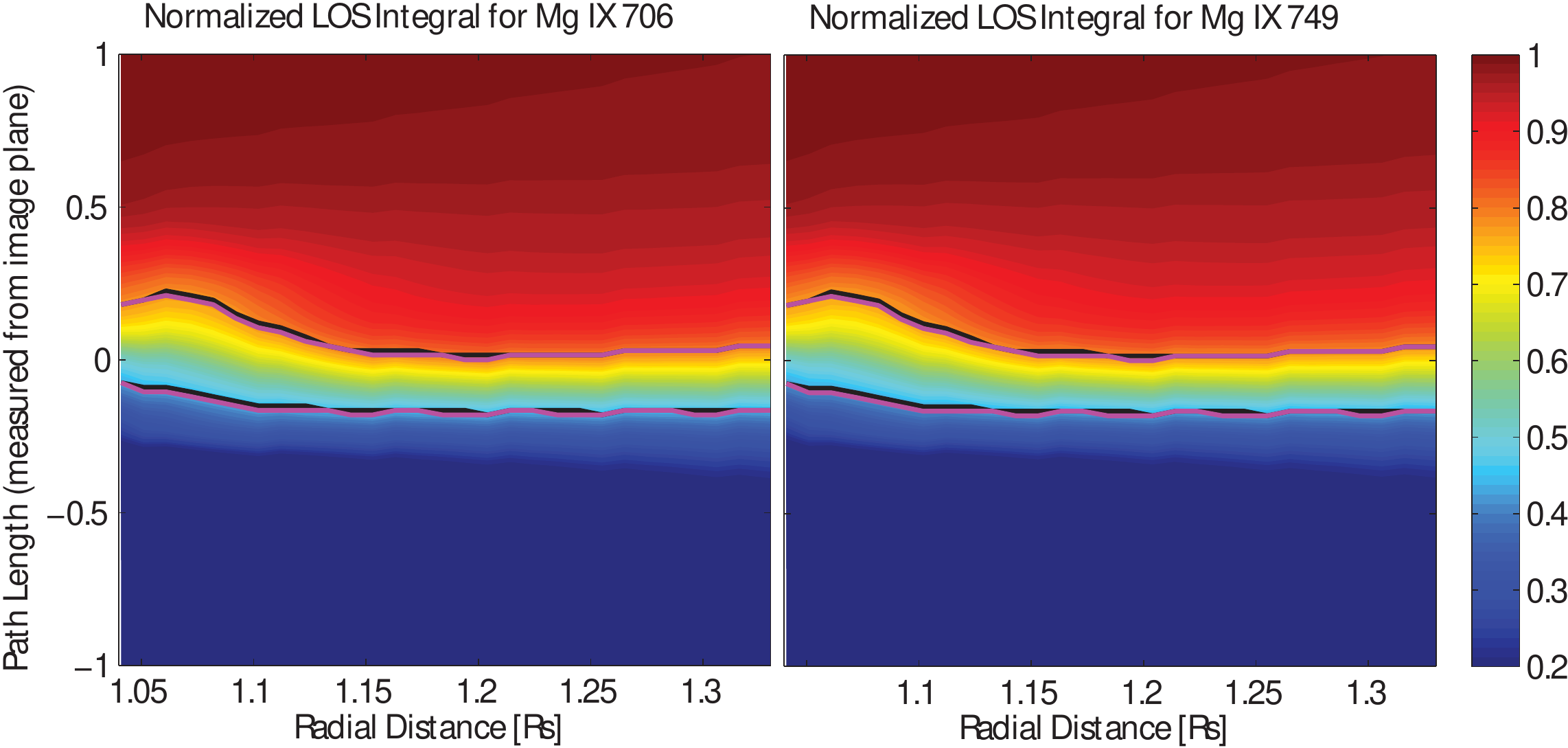} 
\caption{\small \sl Emissivity normalized line of sight integral for  Mg IX 706\AA~ (left) and Mg IX 749\AA~ (right). The blue dots represent the ranges along the LOS that account for 36\% of the total emission of the 706\AA~ line, while the green dots represent the same range for the 749\AA~ line. \label{F:MgIXlines} } 
\end{figure}

\begin{figure}[ht] 
\plotone{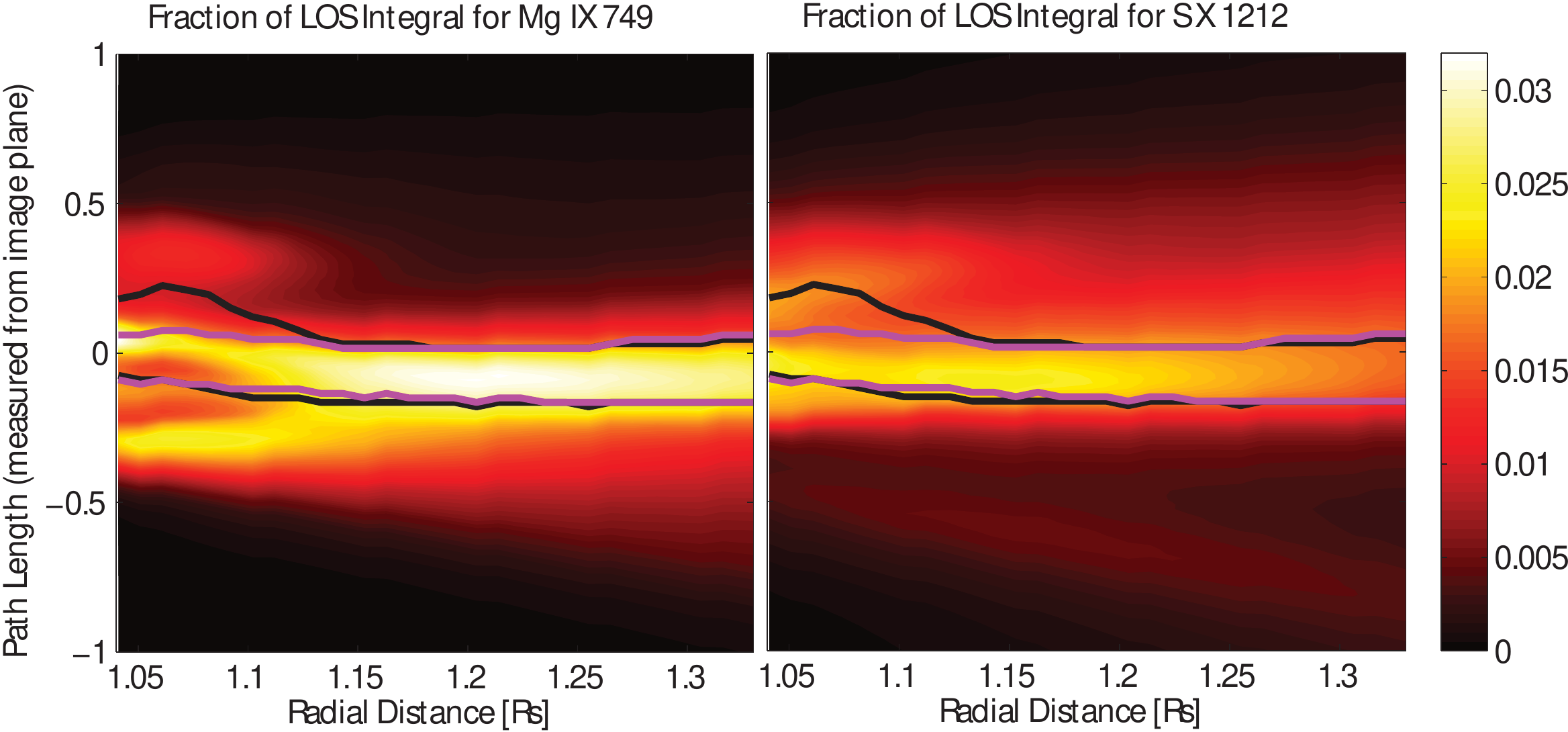} 
\caption{\small \sl Fractional contribution to the line of sight integral for  Mg IX 749\AA~ (left) and S X 1212\AA~ (right). The purple curves represent the ranges along the LOS that the S X line pair has similar contribution (same region as in Figure \ref{F:SXlines}), while the black curves represent the region where the Mg IX line pair has similar contribution (same region as in Figure \ref{F:MgIXlines}).\label{F:MgIX_SX_lines} } 
\end{figure}

The electron temperature along the SUMER slit was obtained from the line flux ratio of Mg IX 706\AA~ and Mg IX 749\AA. As for the S X line pair, we wish to verify that both lines give similar relative contribution to the line-of-sight emission in the pseudo-streamer region. The cumulative contribution along the line of sight is shown in Figure \ref{F:MgIXlines}. The overlaid curves represent the region where the relative contributions of the two lines are similar, and account for 36\% of the total line-of-sight emission. The black and purple curves correspond to the 706\AA~ and the 749\AA~ lines, respectively. As can be seen, these regions almost entirely overlap. Calculating the temperature from the observed line flux ratio also requires us to know the electron density, which we take from the measurement discussed in the previous section. We therefore wish to compare the location of the region of equal contribution of the Mg IX lines to the region of equal contribution of the S X lines, i.e. the pseudo-streamer region selected in the previous section.  The comparison is shown in Figure \ref{F:MgIX_SX_lines}. The panels show the fractional contribution for Mg IX 749\AA~ (left) and for S X 1212\AA~ (right). The purple curves represent the region of equal contribution of the S X line pair (as in Figures \ref{F:SXlines} and \ref{F:S_10_tec}), while the black curves represent the region of equal contribution of the Mg IX line pair (as in Figure \ref{F:MgIXlines}). As can be seen, the spatial distributions of the emission are quite different, mostly at low altitudes. The regions of equal contribution more or less overlap above an heliocentric distance of 1.15 R$_s$. We therefore restrict the comparison of measured and predicted electron temperature to these altitudes only, where we can safely assume that the density and temperature observations apply to the same region. Examining Figure \ref{F:S_10_tec}, we can see that this altitude corresponds to the purely open field line region of the pseudo-streamer, while at lower altitudes the lines of sight intersects both open and closed field line structures.

\subsubsection{Electron Density and Temperature in a Pseudo-Streamer}
We located a distinct and narrow region which accounts for significant and equal parts of the total fluxes used in the electron density and temperature measurements. For each line of sight, we average the predicted quantity over the segment bounded by the white squares in Figure \ref{F:S_10_tec}, to obtain a radial profile along the SUMER slit. \par
\begin{figure}[ht] 
\plotone{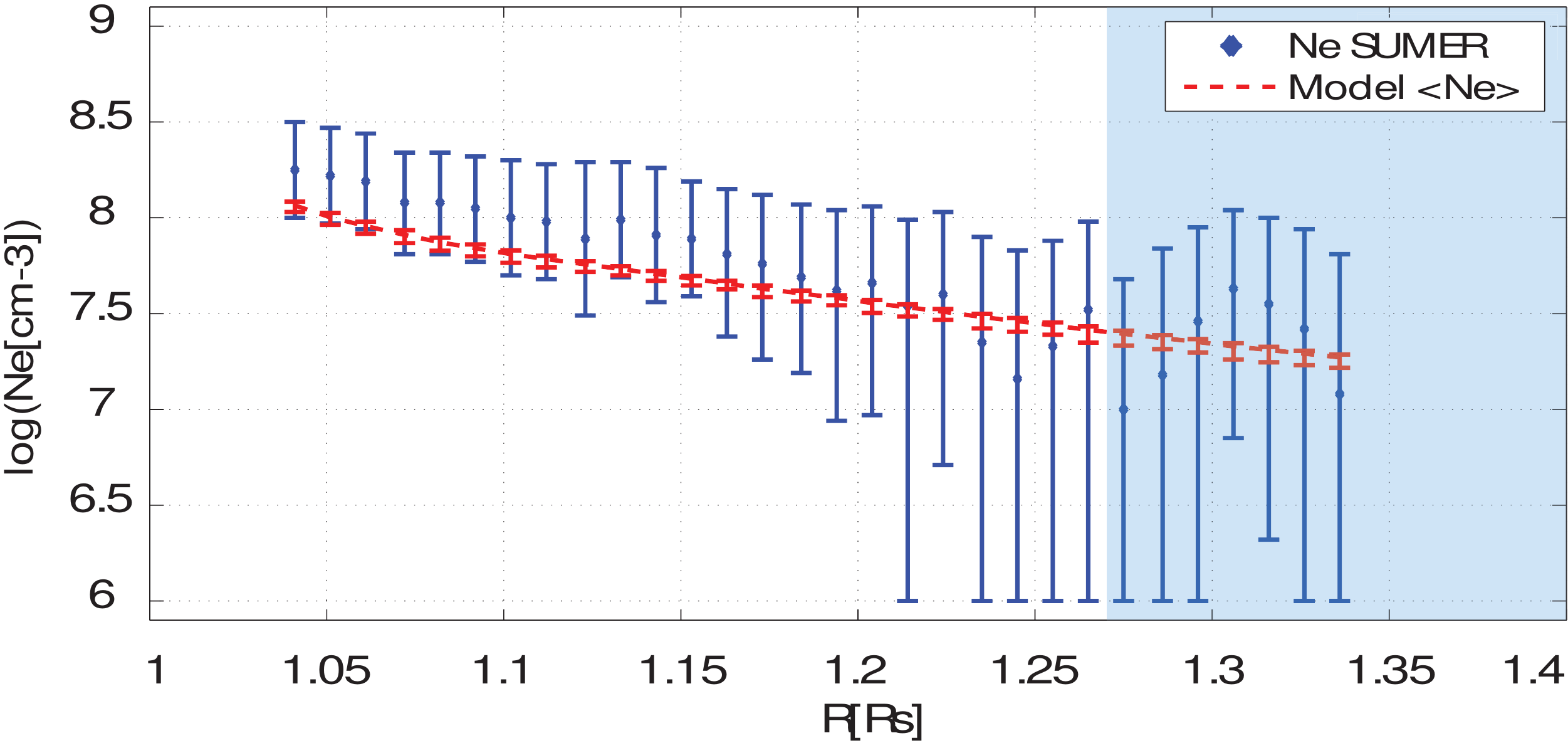} 
\caption{\small \sl Model / SUMER comparison of electron density. The blue curve shows the density measured using the SUMER S X 1196\AA~ and S X 1212\AA~  line flux ratio. The red curve shows the modeled density, averaged over the line of sight segments bounded by white squares in Figure \ref{F:S_10_tec}. The model uncertainty is calculated given the minimum and maximum density along each segment. The shaded region represents the altitude above which the observed line fluxes decreased to below twice the scattered light flux (see Table \ref{T:lines}). \label{F:NePseudo}}
\end{figure}

 
Figure \ref{F:NePseudo} shows the comparison of the predicted electron density in the pseudo streamer with the SUMER measurement. The blue curve with error bars shows the measured electron density while the dashed red line shows the model results. The error bars in the model indicate the minimum and maximum electron density found along the line of sight segments over which we take the average. The shaded region represent the altitude where the observed flux of the lines used for this measurement has decreased to below twice the scattered light flux, making the measurement less reliable at these heights.  As can be seen, the model and measurements are in very good agreement, although the uncertainty in the electron density measurement is quite large.\par
\begin{figure}[ht] 
\epsscale{.6}
\plotone{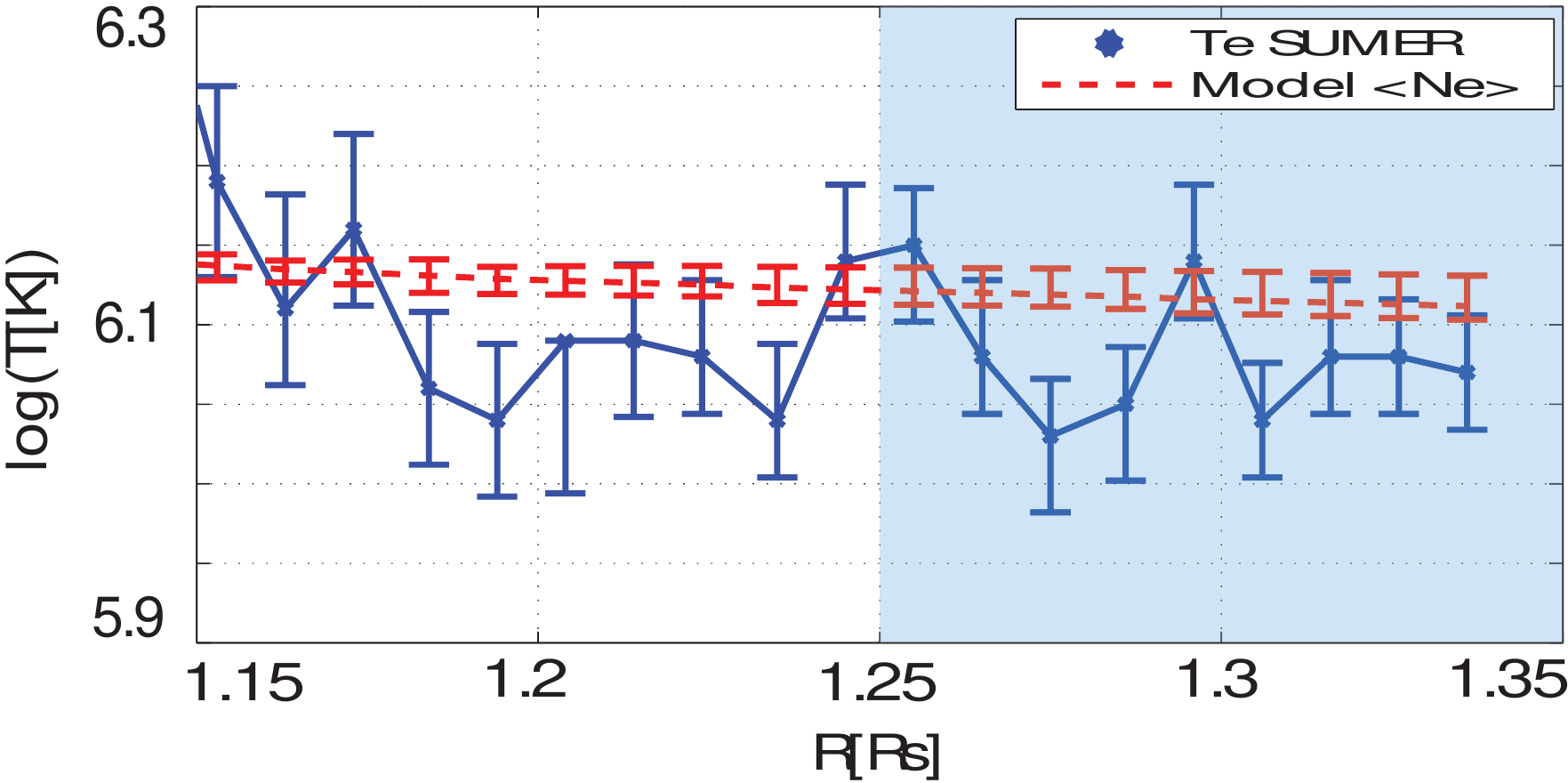} 
\caption{\small \sl Model / SUMER comparison of electron temperature. The blue curve shows the temperature measured using the SUMER Mg IX 706\AA~ and Mg IX 749\AA~ line flux ratio. The red curve shows the modeled electron temperature, averaged over the line of sight segments bounded by white squares in Figure \ref{F:S_10_tec}. The model uncertainty is calculated given the minimum and maximum density along each segment. The shaded region represents the altitude above which the observed line fluxes decreased to below twice the scattered light flux (see Table \ref{T:lines}). \label{F:TePseudo}}
\end{figure}

 
The predicted electron temperature along the SUMER slit and its comparison to observations is shown in Figure \ref{F:TePseudo}. The color coding, as well as the role of the error bars, is the same as in Figure \ref{F:NePseudo}. The comparison starts at $r=1.15$ R$_s$ since below that height the lines used in the temperature measurement are not emitted from the same region as the lines used for the density measurement.  The shaded region corresponds to altitudes where the observed flux of the lines used for this measurement has decreased to below twice the scattered light flux, making the measurement less reliable at these heights.  The measured temperature exhibits large uncertainties and variations with height, with no clear radial trend. The predicted electron temperature falls within the range of observed values, suggesting that the heating supplied by the heating mechanism is sufficient to achieve the observed coronal temperatures in the quiet corona. 
\par

\subsection{Wave Dissipation in the Pseudo-Streamer}\label{S:ducurve}
The three-dimensional, magnetogram-driven solution allows us to study not only the synthetic line of sight line width, but also the variation of the wave amplitude along selected field lines. We recall that the line width observed from a particular direction depends on both the wave energy and the magnetic topology, as is clear from Eq. (\ref{eq:vnt}). Examining Figure \ref{F:S_10_tec}, we can see that the magnetic field in the region of largest emission is composed of a closed loop structure up to a radial distance of $\sim 1.1$R$_s$, above which all field lines are open. In the closed loop region, the magnetic field direction changes from approximately perpendicular to parallel to the line of sight. Thus while the wave amplitude is expected to increase with height in this region (due to the sharp decrease in the plasma density), the fraction of it that lies along the SUMER line of sight will decrease. Above the closed loop structure the magnetic field direction is very close to perpendicular to the line of sight, and thus a larger share of the wave induced motions will contribute to the line width.  This dependence on the line of sight and magnetic field geometry is illustrated in the radial variation of the line widths in Figure \ref{F:all_FWHM}, where an evident change in the synthetic line widths of all ions occurs around $r=1.1$R$_s$. \par
\begin{figure}[ht] 
\epsscale{.9}
\plotone{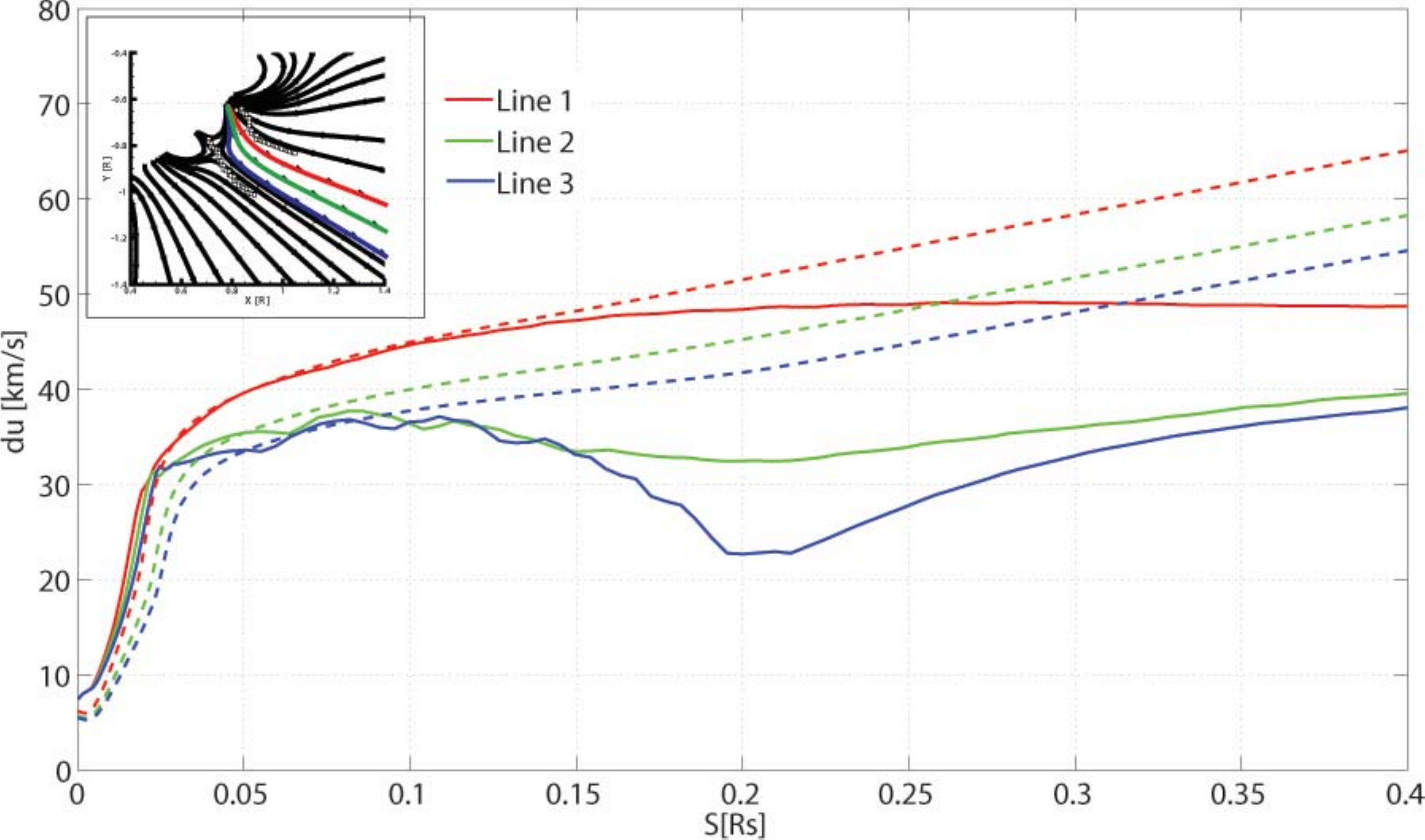} 
\caption{\small \sl Model results of the rms velocity amplitude of Alf{\'e}n waves along selected open field lines. The inset shows the field lines in an equatorial plane. White squares denote the region of maximum emission as described in Section \ref{S:maxemiss}. The solid curves show the rms velocity amplitude extracted from the AWSoM model, while the dashed curves show hypothetical wave amplitudes for undamped waves. The hypothetical curves were normalized to the corresponding modeled value at $S=0.05$R$_s$.  \label{F:VntPseudo}}
\end{figure}

 
In order to study the actual variation of the wave amplitude, we extracted the model results along three open field lines inside the region of largest emission. This will enable us to remove the effects of the line of sight geometry and directly study the wave dissipation taking place in this region. We calculate the rms of the wave velocity amplitude, $\overline{\delta u} = \sqrt{<\delta u^2>}$, using Eq. (\ref{eq:zzave2}). Hassler et al. (1990) and Moran (2001) have shown that if no wave damping is taking place, the rms wave amplitude would vary as $\overline{\delta u}\propto \rho^{-1/4}$ as a result of energy conservation along a magnetic flux tube. Thus we would expect the rms wave amplitude predicted by the model to be lower than the undamped values. The results are shown in Figure \ref{F:VntPseudo}. The location of the selected field lines is shown in the inset. Line 1, colored in blue, is an open field line on the edge of the pseudo-streamer, while line 3, colored in red, in the first open field line straddling the closed loop structure. Line 2, colored in green, lies in between the other two lines. The solid curves show the rms wave amplitude as a function of the path length S along each of the field lines, while the dashed curves show hypothetical curves for undamped waves, normalized to the value of the modeled curve at $S=0.05$R$_s$. As expected, the rms wave amplitude sharply increases close to the inner boundary due to the sharp drop in density. Departures from the undamped curve become prominent above $S=0.05-0.1$R$_s$, although each of the field lines exhibit a different dissipation rate. It is interesting to compare the damped and undamped curves to a similar analysis presented in \citet{Hahn2012} for a polar coronal hole. In Figure 5 therein, the observed effective speeds of several emission lines are compared to undamped values. Departures from the undamped curves start above heights of $0.1$R$_s$ and $0.2$R$_s$ above the limb, depending on the ion. Of the three field lines in Figure \ref{F:VntPseudo}, line 1 most resembles a coronal hole field line, with minimal bending around the closed loop structure. The wave amplitude along line 1 shows very similar behavior to the one reported in \citet{Hahn2012}. In the case of line 2 and line 3, larger departures from energy conservation occur at lower heights near the closed loop region. This is most prominent for line 3, where the wave amplitude is significantly reduced near the tip of the loop structure. In this location, higher dissipation is expected to occur due to the presence of counter-propagating waves, and the first term under the square root in Eq. (\ref{eq:dissipfinal}) will be taken into account. Above that point, the rms wave amplitude increases at a rate similar to that of line 1, consistent with the fact that the dissipation rate is now dominated by reflections, i.e. the second term under the square root in Eq. (\ref{eq:dissipfinal}). Line 2 also exhibits a signature of this behavior, although it is less pronounced.

\section{Conclusions}\label{S:discussion}
In this work, we have examined whether the dissipation of Alfv{\'e}n waves due to a turbulent cascade is a likely candidate to explain the observed large-scale distribution of coronal heating rates. By combining results from an Alfv{\'e}n wave-driven MHD model with the CHIANTI atomic database, we were able to produce, for the first time, synthetic EUV spectra that include thermal and non-thermal broadening from a global model. \par 
The ability to predict non-thermal line broadening in a wave-driven global model is an important step in testing the validity of the underlying wave heating mechanism, as this observable is directly related to wave-induced motions and is a measure of the modeled amplitude of the Alfv{\'e}n waves. The advantage of a global model is that the predicted emission is integrated over the line of sight using the full three-dimensional solution, without invoking simplifying assumption about the geometry of the system. \par 
Comparing the synthetic spectra to detailed SUMER observation between r=1.03  - 1.43R$_s$, we tested whether the AWSoM model can predict plasma properties and wave energies that are simultaneously consistent with observations. The predicted total flux in selected emission lines depends on the electron density and temperature, while the line width depends on the ion temperature and wave amplitude. We have found good agreement between predicted and observed line width, and reasonable agreement of the total flux, given the uncertainties in atomic data.\par 
By taking advantage of the three-dimensional nature of the solution, we could calculate the relative contribution of different regions along the line of sight to the observed emission. We found that a substantial fraction of the emission of several lines comes from a narrow, well defined magnetic structure: an equatorial pseudo-streamer. The electron density and temperature predicted by the model are in good agreement with the measurements performed using the emission of these lines, suggesting that this region is indeed the source of the relevant radiation detected by SUMER. This type of three-dimensional line of sight analysis is important to the interpretation of any remote observation.\par 
In summary, we have shown that the treatment of Alfv{\'e}nic energy as described in the AWSoM model simultaneously produces electron densities, temperatures, total line fluxes and line broadening that are consistent with observations. This suggests that the model correctly describes the amount of wave energy injected into the system, and the fraction of it that is deposited as heat.\par 
Finally, we mention possible improvements and future work. First, the synthetic profiles can be calculated more accurately. The line fluxes calculations used here were based on the assumption of ionization equilibrium. This assumption may break down, as wind-induced departures from equilibrium may occur. A more accurate calculation should be based on solving the charge state evolution in the region under question, which will be the basis of a more accurate calculation of the line fluxes. Second, the model's treatment of wave propagation and dissipation can be improved. Most notably, our treatment of wave reflections is not done self-consistently, as the reflection coefficient should depend on the magnetic topology. This requires a much more complex treatment of the wave field and its coupling to the MHD plasma. Such a treatment is presented in {vanderholst2013}


\acknowledgments
\begin{center}
\bf{Acknowledgments}\normalfont\\
\end{center}
This work was supported by the NSF grant AGS-1322543 (Strategic Capabilities).
The work of E. Landi is supported by NASA grants NNX10AQ58G, NNX11AC20G, and 
NNX13AG22G. The simulations performed in this work were made possible thanks to the NASA Advanced Supercomputing Division, which granted us access to the Pleiades Supercomputing cluster.
Analysis of radiative processes was made possible through the use of the CHIANTI atomic database. CHIANTI is a collaborative project involving the following Universities: Cambridge (UK), George Mason and Michigan (USA).\par


\end{document}